\documentclass[journal]{IEEEtran}
\usepackage[font={small}]{caption}
\usepackage[table]{xcolor}
\usepackage{amsmath,graphicx}
\usepackage{algorithm}
\usepackage{algorithmicx}
\usepackage{algpseudocode} 
\usepackage{array}
\usepackage{subfig}
\usepackage{textcomp}
\usepackage{stfloats}
\usepackage{siunitx}
\usepackage{url}
\usepackage{siunitx}
\usepackage{verbatim}
\usepackage{graphicx}
\usepackage{mathtools}

\usepackage{mathrsfs}
\usepackage{bm}
\usepackage{amssymb}
\usepackage{multirow}
\usepackage{booktabs}
\usepackage{threeparttable} 
\usepackage[numbers,sort&compress]{natbib}
\begin{document} 
\title{SemHARQ: Semantic-Aware Hybrid Automatic Repeat Request for Multi-Task Semantic Communications}

\author{Jiangjing~Hu, Fengyu~Wang{\textsuperscript{$\ast$},~\IEEEmembership{Member,~IEEE}},
        Wenjun~Xu,~\IEEEmembership{Senior Member,~IEEE},
        Hui~Gao,~\IEEEmembership{Senior Member,~IEEE}, and Ping~Zhang,~\IEEEmembership{Fellow,~IEEE}
        % <-this % stops a space
\thanks{This work was supported in part by the National Natural Science Foundation of China under Grant 62301069 and 62293485, and in part by the Fundamental Research Funds for the Central Universities under Grant 2025TSQY11~ \emph{(Corresponding author: Fengyu Wang)}.}% <-this % stops a space
\thanks{Jiangjing Hu, Wenjun Xu and Ping Zhang are with the State Key Laboratory of Network and Switching Technology, Beijing University of Posts and Telecommunications, Beijing 100876, China~(e-mail: hujiangjing@bupt.edu.cn, wjxu@bupt.edu.cn, pzhang@bupt.edu.cn). 

Fengyu Wang is with the School of Artificial Intelligence, Beijing University of Posts and Telecommunications, Beijing 100876, China~(e-mail: fengyu.wang@bupt.edu.cn).

Hui Gao is with the Key Laboratory of Trustworthy Distributed
Computing and Service, Ministry of Education, Beijing University of Posts and Telecommunications, Beijing 100876, China~(e-mail: huigao@bupt.edu.cn).}}

% Ping Zhang is with the State Key Laboratory of Networking and Switching Technology, Beijing University of Posts and Telecommunications, Beijing 100876, China, and also with Peng Cheng Laboratory,
% Shenzhen 518066, China (email: pzhang@bupt.edu.cn).

% The paper headers
% \markboth{Journal of \LaTeX\ Class Files,~Vol.~14, No.~8, August~2021}
% {Shell \MakeLowercase{\textit{et al.}}: A Sample Article Using IEEEtran.cls for IEEE Journals}

%\IEEEpubid{0000--0000/00\$00.00~\copyright~2021 IEEE}
% Remember, if you use this you must call \IEEEpubidadjcol in the second
% column for its text to clear the IEEEpubid mark.

\maketitle
\begin{abstract}
Intelligent task-oriented semantic communications~(SemComs) have witnessed great progress with the development of deep learning~(DL), where multi-task SemComs that perform multiple tasks simultaneously attach great importance due to its high efficiency. However, the study of robust multi-task-oriented semantics transmission is still in early stages. In this paper, we propose a semantic-aware hybrid automatic repeat request~(SemHARQ) framework for the robust and efficient transmissions of multi-task semantic features. First, to improve the robustness and effectiveness of semantic coding, a multi-task semantic encoder is proposed. Meanwhile, a feature importance ranking~(FIR) method is investigated to ensure the important features delivery under limited channel resources. Then, to accurately detect the possible transmission errors, a novel feature distortion evaluation~(FDE) network is designed to identify the distortion level of each feature, based on which an efficient HARQ method is proposed. Specifically, the corrupted features are retransmitted, where the remaining channel resources are used for incremental transmissions. The system performance is evaluated under different channel conditions in multi-task scenarios in the Internet of Vehicles. Extensive experiments show that the proposed framework outperforms state-of-the-art works by more than $20\%$ in rank-1 accuracy for vehicle re-identification, and $10\%$ in vehicle color classification accuracy in the low signal-to-noise ratio regime.
\end{abstract}


\section{Introduction}

\IEEEPARstart{I}n {the era of the sixth-generation~(6G) wireless communications, communications among humans, machines, objects, and intelligent agents enable intelligent services to individuals and industries~\cite{6g,6g2,6g3}. As a result, the requirements for communication systems to effectively convey the intended meaning have increased.}  Conventional communication systems focus on the successful transmissions of bits but ignore the meanings of messages, facing critical challenges to support the aforementioned applications over limited communication resources~\cite{ref1,weaver1953recent}. To solve this issue, semantic communications~(SemComs), which transmit the intention-related semantics, can significantly reduce the amount of transmission data while ensuring the system performance. Consequently, the efficiency of communication is improved effectively. {Meanwhile, with the development of artificial intelligence (AI), deep learning~(DL) has shown its superiority in semantic representation~\cite{sem-represen}, compression~\cite{sem-compre} and recovery~\cite{sem-recovery}, which has been widely used in task-oriented SemComs to enable the extraction, compression, and reconstruction of semantic features~\cite{task-oriented-jsac,image-retrival}.}

To guarantee the reliability of semantic transmissions, most of the current DL-based SemCom systems transmit task-related semantics and identify semantic errors by joint source-channel coding~(JSCC)~\cite{jscc2,jscc3}. Particularly, Bourtsoulatze $\emph{et}$ $\emph{al}$. propose a JSCC technique for image transmission, which directly maps the image pixel values to the complex-valued channel input symbols~\cite{jscc2}. Furthermore, besides transmitting semantics in an end-to-end~(E2E) manner, a multi-user cooperative semantic framework based on JSCC is proposed to exploit the correlations among users in~\cite{jscc3}. However, the above SemCom systems are designed with a fixed coding rate, incapable of transmitting semantics adaptively according to the dynamic channel conditions. 
To overcome this problem, a variable length coding method is proposed based on dynamic neural networks in~\cite{IB-jsac}. Meanwhile, a policy network is trained additionally in~\cite{IB-harq} to select {a part of} the encoded feature vector based on the signal-to-noise ratio~(SNR). Nevertheless, the robustness of semantic recovery is limited {in the
mentioned systems due to the lack of an accurate processing on semantic errors}.

To improve the robustness of communication systems, hybrid automatic repeat request~(HARQ) is commonly used in conventional communications~\cite{tran-harq}. {Specifically, the received information packet is detected by a cyclic redundancy check~(CRC) detector at the receiver. Then, a NACK signal is sent back to the transmitter to trigger a retransmission if the detection fails. The process will repeat until the packet is received correctly, or the maximum number of transmissions is reached. Particularly, based on the type of retransmitted data, the conventional HARQ can be classified into three main types. In type-I HARQ~(HARQ-I), the same data packet is retransmitted, and the erroneous packet is replaced with the new one~\cite{harqI}. Since the erroneous packets are simply discarded in HARQ-I, the useful information is not fully utilized, reducing the efficiency of retransmissions. To overcome this problem, in type-II HARQ~(HARQ-II), the original bit vector is punctured into different lengths, and the incremental redundancy is transmitted when retransmission is needed. Then, at the receiver, the received packet is combined with the previously transmitted information for the forward error correction~(FEC) decoding~\cite{harqII}. However, due to the lack of information bits, the received incremental redundancy cannot be decoded independently, degrading the efficiency of decoding. Therefore, type-III HARQ~(HARQ-III) is proposed to include the information bits and parity bits simultaneously at every
retransmission~\cite{harqIII}.} {Nevertheless, the meanings of transmitted information are ignored in the above methods, making it incapable of handling semantic errors. Moreover, since the CRC detector can only detect whether the entire received packet is erroneous, the fine-grained error detection is not considered, degrading the efficiency of error detection.}

Therefore, to further enhance the robustness and efficiency of SemComs, semantic retransmission schemes have been studied~\cite{retransmission1,retransmission2,retransmission3,retransmission4,retransmission5,zhengharq}. In~\cite{retransmission1}, the data sample is thoroughly retransmitted in a DL-based communication network when the received semantic information is not sufficient to complete the task. A channel output feedback is further incorporated into the retransmission mechanism in~\cite{retransmission2} to improve the reconstruction quality. Although constantly retransmitting messages can improve the performance of specific intelligent task, the efficiency in resource utilization is too low for practical usage. Therefore, incremental transmission methods are further incorporated to improve the transmission efficiency. Specifically, multiple incremental transmissions are supported by different autoencoders in~\cite{ir-harq2} to reduce semantic transmission error. To further simplify the network architecture, a unified single decoder is proposed in~\cite{IB-harq} to exploit the incremental knowledge~(IK) obtained from retransmission. Besides the text and image transmission, an incremental redundancy HARQ framework is investigated for video conferencing with a semantic error detector~\cite{ir-harq3}. However, the above works retransmit the same messages simply or directly transmit incremental redundancy triggered by task performance, having no insight into the specific distortion from the channel fading and noise. Consequently, the error detection on the semantic feature dimension is always ignored, where the features that are successfully received and decoded are probable to be transmitted again, decreasing the efficiency of resource utilization. At the same time, the receiver may fail to recover critical semantics that are previously transmitted when only incremental transmissions are involved without considering retransmissions, degrading the system performance.

Moreover, to ensure the transmission of critical semantics with a higher priority, the impact of different importance levels of semantic features has been investigated~\cite{importance0,importance1,importance2,importance3}. Wang $\emph{et}$ $\emph{al}$. evaluate the importance of each semantic triple using an attention network~\cite{importance1,importance2}. Meanwhile, the importance diversity among semantic features is estimated in~\cite{importance3} from an information theory perspective, where an entropy model is set on the latent space. Nevertheless, the above measurements cannot be associated with the semantic task performance directly, degrading the task efficiency.

Overall, to realize robust and scalable semantic transmissions, in this paper, we propose a semantic-aware HARQ~(SemHARQ) framework for multi-task semantic communications. The major contributions of this work are summarized as follows:

\begin{itemize}
		\item[$\bullet$] {First, to avoid unnecessary retransmissions, a \emph{{feature distortion evaluation}}~(FDE) network is proposed to differentiate the distortion extents of the received features.  The corrupted features are then identified and retransmitted, and meanwhile, the remaining resources are utilized to transmit the incremental semantics for better performance. }
		\item[$\bullet$] {Then, to further improve the transmission efficiency, a \emph{{feature importance ranking}}~(FIR) method is designed for multi-task SemComs. Specifically, a multi-task semantic encoder is proposed~\cite{multi-task1,multi-task2,multi-task4}, where the correlated semantics among different tasks are extracted jointly, making the semantic coding more robust and efficient than existing single-task coding~\cite{single3,single4}. Moreover, the semantic importance levels are further evaluated by measuring the sensitivity of task performance to different features, where  the gradients of the task performance {with respect to} the features are considered~\cite{importance4}.
                          }
        \item[$\bullet$]
        {
        Finally, to enhance the scalability of the proposed SemHARQ, the amount of semantics at each transmission is adaptively adjusted. Specifically, the length of the semantic feature vector to be transmitted is determined by the channel conditions. During retransmission, besides the features that failed to be reconstructed in previous transmissions, the incremental features are also transmitted in descending order of importance to fully utilize channel resources.
        }
        \item[$\bullet$]
        {
        Extensive simulations are conducted to validate the effectiveness of the proposed SemHARQ. Numerical results demonstrate that SemHARQ outperforms the traditional HARQ and state-of-the-art semantic-aware HARQ methods, especially at the low SNR regime. To the best of our knowledge, this is the first HARQ method that realizes semantic feature-level retransmission considering semantic importance evaluation.
        }
                          
\end{itemize}

The remainder of this paper is organized as follows. Section II introduces the system model of the proposed SemHARQ. The architecture and principle of the FDE network and FIR method are illustrated in Section III. In Section IV, the process and training strategy of the SemHARQ are presented.  Numerical results of the proposed system are demonstrated in Section V. Finally, Section VI concludes this paper.

{\emph{Notation:} $\mathbb{C}^{n}$ and $\mathbb{R}^{n}$ represent the set of $n$-dimensional complex and real vectors, respectively. Bold-font variables denote vectors, where $x_i$ represents the $i$-th value of vector $\bm{x}$. Additionally, note that in this paper, the superscript of a vector indicates the number of retransmissions. For example, $\bm{x}^j$ indicates the vector corresponding to the $j^{th}$ retransmission. Finally, $x \sim \mathcal{CN}(\mu, \sigma^2)$ means variable $x$ follows a circularly-symmetric complex Gaussian distribution with mean $\mu$ and covariance $\sigma^2$. }

\begin{figure}[t]
  \centering
  \centerline{\includegraphics[width=8cm]{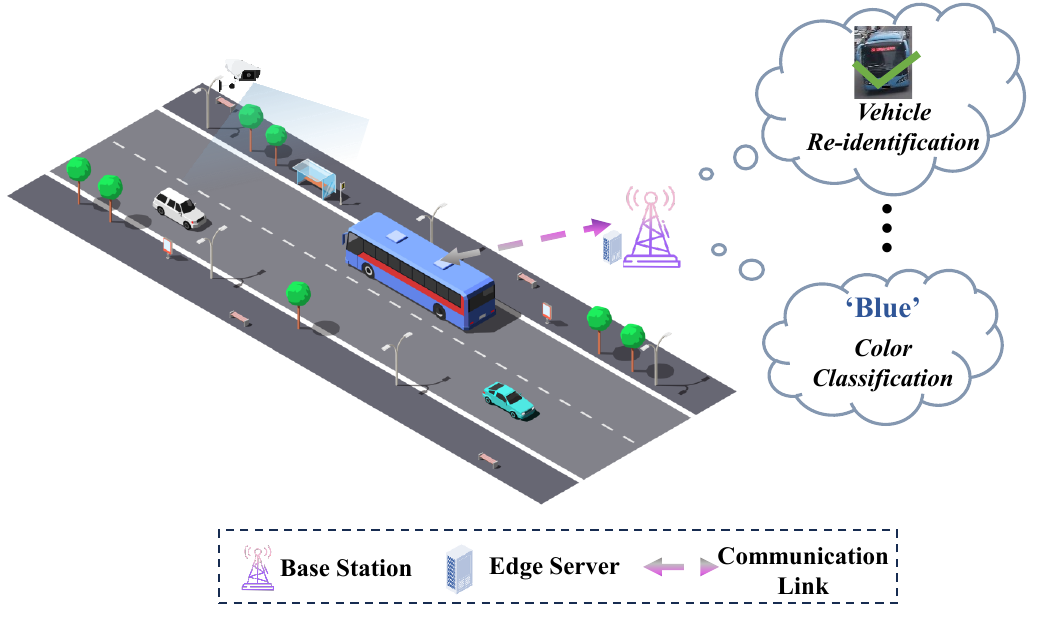}}
%  \vspace{2.0cm}
  \captionsetup{justification=centering}
  \caption{The scenario of the multi-task intelligent transportation systems.}
  \label{fig:Scenario}
\end{figure}

\begin{figure*}[t]
  \centering
  \centerline{\includegraphics[width=15cm]{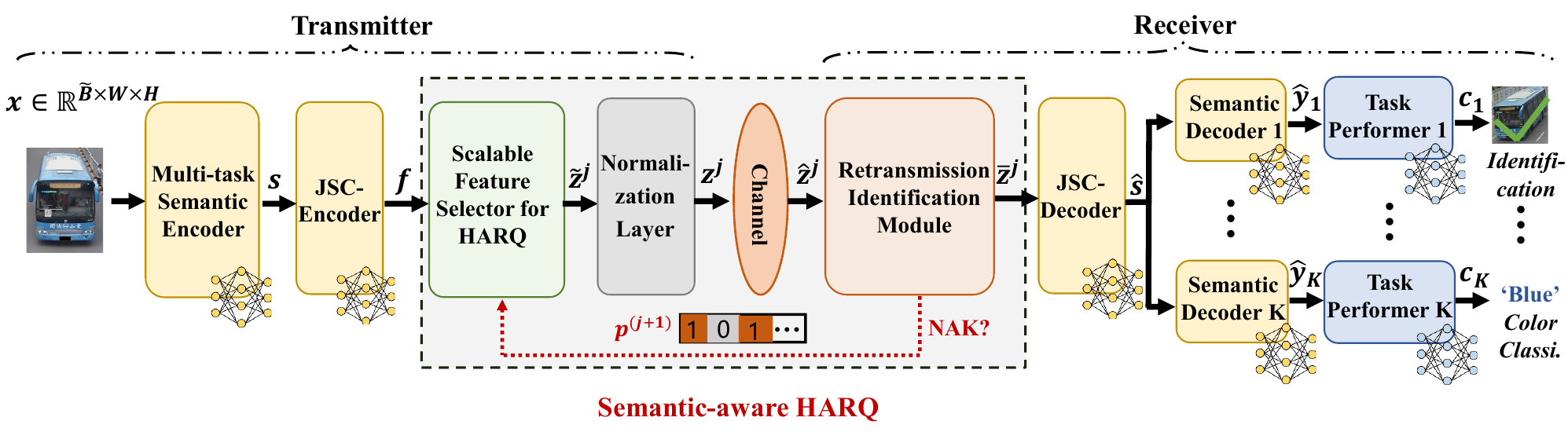}}
%  \vspace{2.0cm}
  \captionsetup{justification=centering}
  \caption{The framework of the proposed SemHARQ.}
  \label{fig:Semharq framework}
\end{figure*}

\section{System Model}
In this section, the system model of SemHARQ is introduced. Specifically, $K$ intelligent tasks are executed simultaneously in intelligent transportation scenarios, and the inputs of the system are assumed to be vehicle images, as shown in Fig.~\ref{fig:Scenario}. Note that the system can be readily extended to other scenarios and modalities. {First, the semantic features are extracted and encoded as symbols, then directly transmitted to the receiver after mapping into complex symbols and scalably selected according to the importance levels. Then, the SemHARQ is triggered after receiving the semantic features from the channel.  
After all the transmissions, the concatenated feature vector is decoded by a joint source-channel decoder~(JSC-Decoder) and multiple semantic decoders. Finally, the multiple tasks are performed by different task performers.}

%\subsection{Transmitter} 
As shown in Fig.~\ref{fig:Semharq framework}, the input images  denoted as $\bm{x}\in\mathbb{R}^{\widetilde{B} \times W \times H}$ are encoded by a multi-task semantic encoder, where $\widetilde{B}$ is the batch size, and $W$, $H$ are the width and height of an image, respectively. The semantic features are extracted as $\bm{s}\in\mathbb{R}^N$, where $N$ is the number of features in $\bm{s}$. Then, a JSC-Encoder encodes $\bm{s}$ as $\bm{f}\in\mathbb{C}^L$, where $L$ is the length of the feature vector. Afterwards, the feature vector selected by a scalable feature selector for HARQ is represented as ${\bm{\widetilde{z}}}^j\in\mathbb{C}^B$, where $B$ is the number of features to be transmitted according to channel state information~(CSI), $j$ is the number of retransmissions. The value of $B$ is determined by the transmission bandwidth in a single time slot so that the length of the transmitted feature vector can be scalably adjusted according to the channel conditions. Finally, the average power control is utilized to normalize the input signals of the channel at each transmission, which is shown as
\begin{equation}
\label{deqn_ex1}
{\bm{z}}^j= \sqrt{PB}\frac{{\bm{\widetilde{z}}}^j}{||{\bm{\widetilde{z}}}^j||_2},
\end{equation}
where $P$ is the average power. 

Furthermore, ${\bm{z}}^j$ is sent and passes through the stochastic physical channel. The channel output can be represented as 
\begin{equation}
\label{deqn_ex1}
{\bm{\hat{z}}}^j=\bm{h}{\bm{z}}^j+\bm{n},
\end{equation}
where $\bm{n}\in \mathbb{C}^B$ denotes the white Gaussian noise, the value of which follows ${n}_i\sim\mathcal{C}\mathcal{N}(0,\sigma^2)$. $\bm{h}\in \mathbb{C}^B$ represents the fading coefficient of the channel. In this paper, {the performance is evaluated in the intelligent transportation scenario as shown in Fig.~\ref{fig:Scenario}, where both the line-of-sight~(LOS) and the non-line-of-sight~(NLOS) paths should be considered~\cite{iov-rician1,iov-rician2}. As a result, the Rician channel is considered}.  The fading coefficient follows $\mathcal{C}\mathcal{N}\Bigl(\sqrt{r/(r+1)},1/(r+1)\Bigl)$, where $r$ is set to $2$.

%\subsection{Receiver} 
At the receiver, ${\bm{\hat{z}}}^j$ is measured by a retransmission identification module to detect if the retransmission is required. If  ${\bm{\hat{z}}}^j$ passes the detection, it will be directly concatenated with the previously received features and the concatenated vector is denoted as ${\bm{\overline{z}}}^j=[{\bm{\hat{z}}}^1,...,{\bm{\hat{z}}}^j]$. Otherwise,  the received ${\bm{\hat{z}}}^j$ will be evaluated by the feature distortion evaluation~(FDE) network to generate a binary feedback vector {${\bm{p}}^{(j+1)}$}, indicating which features in  ${\bm{\hat{z}}}^j$ are needed to be retransmitted. Such retransmissions will be repeated until the success of multi-task or the maximum times of retransmission is reached.  After all the transmissions, the concatenated vector is decoded as $\bm{\hat{s}}\in\mathbb{R}^N$ by the JSC-Decoder. $\bm{\hat{s}}$ {is then semantically decoded} to $[\bm{\hat{y}}_1,\bm{\hat{y}}_2,...,\bm{\hat{y}}_K]$ by multiple semantic decoders, where $K$ represents the number of tasks. Finally, the task prediction vectors $[\bm{c}_1,\bm{c}_2,...,\bm{c}_K]$ are generated from the task performers.

\section{Feature Distortion Evaluation and Importance Ranking}

In this section, the FDE network and feature importance ranking~(FIR) method are introduced. Firstly, the FDE network is investigated to locate the corrupted features that need to be retransmitted in the received feature vector. Then, the FIR method is designed to identify important features in the scalable feature selector for HARQ.

\subsection{FDE network}
\label{FDE}

\begin{figure}[t]
  \centering
  \centerline{\includegraphics[width=7cm]{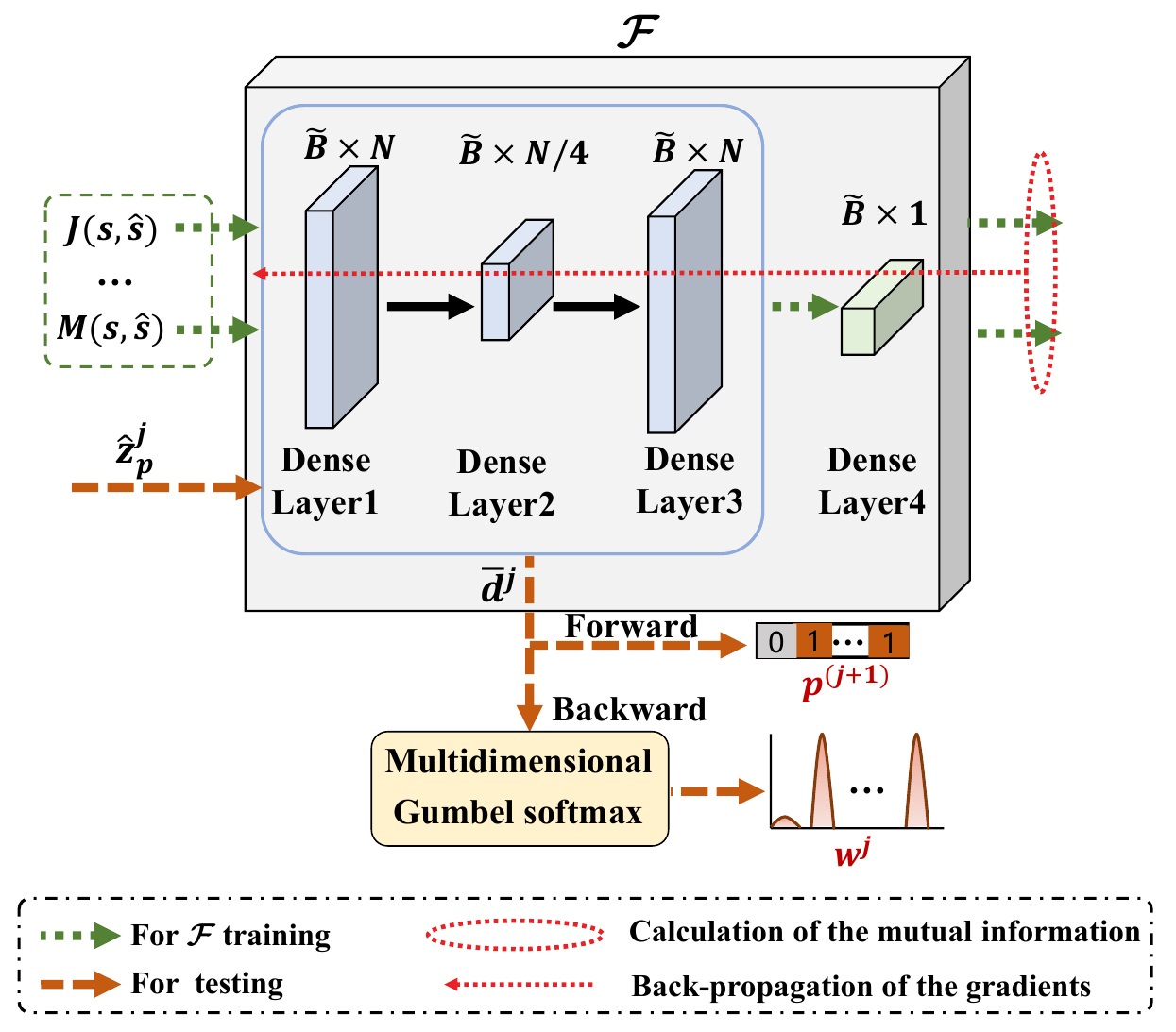}}
%  \vspace{2.0cm}
\caption{{The structure of the FDE network.}}
\captionsetup[figure,justification=raggedright]{labelsep=space}
\label{fig2}
\end{figure}

{As shown in Fig.~\ref{fig2}, to obtain the distortion extents of the received features, a distortion measure module $\mathcal{F}$ is proposed. Specifically, a neural network~(NN)  consisting of four dense layers is utilized to evaluate the distortion distribution of the received vector ${\bm{\hat{z}}}^j$ after $j^{th}$ retransmission. {Since ${\bm{\hat{z}}}^j$ corresponds to a subset of the original features after scalable feature selection, to align the features in ${\bm{\hat{z}}}^j$ with the original feature vector $\bm{s}$, a vector ${\bm{\hat{z}}}^j_\mathrm{p}$ of the same dimension as the original vector is generated from ${\bm{\hat{z}}}^j$ by padding the other positions with zeros. Then, a distortion measure vector can be generated by feeding ${\bm{\hat{z}}}^j_\mathrm{p}$ into $\mathcal{F}$, which has the same dimension as ${\bm{\hat{z}}}^j_\mathrm{p}$, inferring the distortion levels of all the features in ${\bm{\hat{z}}}^j$.}}

{To enable $\mathcal{F}$ to learn the discrepancy of the distributions between the received features corrupted by the channel impairments and the original undistorted feature vector $\bm{s}$, $\mathcal{F}$ is trained independently before the entire system training. Since mutual information can be used to measure the semantic similarity between the transmitted and received features, the mutual information of the original feature vector $\bm{s}$ and the JSC-decoded feature vector $\bm{\hat{s}}$ is estimated and serves as the training objective of $\mathcal{F}$.}

{Specifically, the optimization problem of FDE network training is defined as}
\begin{equation}
    \max_{{\mathcal{F}}} \;\; I(\bm{s};\bm{\hat{s}}).
    \label{eq:IM optimization}
\end{equation}
The mutual information of $\bm{s}$ and  $\bm{\hat{s}}$ can be calculated as~{\cite{elements}},
\begin{equation}
\label{KL}
I(\bm{s};\bm{\hat{s}})={\mathbb{E}}_{p(s,\hat{s})}\bigg[\log\frac{p(s,\hat{s})}{p(s)p(\hat{s})}\bigg]=D_{\rm KL}\Bigl(p(s,\hat{s})||p(s)p(\hat{s})\Bigl),
\end{equation}
where $p(s)$ and $p(\hat{s})$ are the marginal probability of $\bm{s}$ and $\bm{\hat{s}}$, respectively, and $p(s,\hat{s})$ is the joint probability of $\bm{s}$ and $\bm{\hat{s}}$. Note that the mutual information can be represented by the Kullback-Leibler (KL) divergence between the product of the marginal probabilities and the joint probability.
Inspired by the general-purpose parametric mutual information neural estimation~(MINE) method of ~\cite{mi}, the KL-divergence in Eq.~(\ref{KL}) obeys the following dual representation [\citenum{mi}, Eq.~(5)],
\begin{equation}
\label{sup}
    \begin{aligned}
D_{\rm KL}\Bigl(p(s,\hat{s})||p(s)p(\hat{s})\Bigl)=\\\sup _{\mathcal{T}: \Omega \rightarrow \mathbb{R}} \mathbb{E}_{p(s,\hat{s})}[\mathcal{T}]-\log \left(\mathbb{E}_{p(s)p(\hat{s})}\left[e^{\mathcal{T}}\right]\right),
    \end{aligned}
\end{equation}
where the supremum is taken over all functions $\mathcal{T}$. Consequently, the lower bound of $I(\bm{s};\bm{\hat{s}})$ can be derived {[\citenum{mi}, Eq.~(6)]},
\begin{equation}\label{lower bound}
\begin{aligned}
I(\bm{s};\bm{\hat{s}}) &= D_{\rm KL}\Bigl(p(s,\hat{s})||p(s)p(\hat{s})\Bigl) \\
&\geqslant {\mathbb E}_{p\left( {s,\hat{s}} \right)}\left[ \mathcal{T} \right] - \log \left( {{{\mathbb E}_{p\left( s \right)p\left( \hat{s} \right)}}[{e^\mathcal{T}}]} \right).
\end{aligned}
\end{equation}
Note that $\mathcal{T}$ can be any class of functions satisfying integrability, as a result, the NN in $\mathcal{F}$ is employed as $\mathcal{T}$ to obtain a precise bound according to Eq.~(\ref{lower bound}). Finally, the estimated mutual information can be denoted as
\begin{equation}\label{estimated mi}
I(\bm{s};\bm{\hat{s}}|\mathcal{F}) = {{\mathbb E}_{p\left( {s,\hat{s}} \right)}}\left[ \mathcal{F} \right] - \log \left( {{{\mathbb E}_{p\left( s \right)p\left( \hat{s} \right)}}\left[ {{e^{\mathcal{F}}}} \right]} \right),
\end{equation}
where  $\mathcal{F}$ is trained in an unsupervised manner and the optimization goal is updated to maximize $I(\bm{s};\bm{\hat{s}}|\mathcal{F})$~(\textcolor{black}{i.e., $-I(\bm{s};\bm{\hat{s}}|\mathcal{F})$ serves as a part of $L_{\text{FDE}}$, which is the loss of FDE network training}). {As indicated by Eq.~(\ref{estimated mi}), during the training phase of $\mathcal{F}$, the network should be optimized using both joint and marginal distributions of $\bm{s}$ and $\bm{\hat{s}}$. Thus, as shown in Fig.~\ref{fig2}, the inputs are defined as $J(\bm{s},\bm{\hat{s}})$ and $M(\bm{s},\bm{\hat{s}})$, which are sampled from the joint feature space and the marginal feature space of $\bm{s}$ and $\bm{\hat{s}}$, respectively. Particularly, $J(\bm{s},\bm{\hat{s}})$ are sampled from the combination of $\bm{s}$ and $\bm{\hat{s}}$ corresponding to the same images, and $M(\bm{s},\bm{\hat{s}})$ are sampled from the combination of $\bm{s}$ and $\bm{\hat{s}}$ corresponding to different images within a batch. As a result, $\mathcal{F}$ can learn the distributions of $\bm{s}$ and $\bm{\hat{s}}$.} 

{ By optimizing the parameters through maximizing the mutual information, the first three dense layers in $\mathcal{F}$ learn to make the received feature vector approach the distributions of the original undistorted vectors across all the categories in the dataset. On this basis, the output vector $\Bar{\bm{d}}^j$ as shown in Fig.~\ref{fig2} can serve as a weight vector that prominently quantifies the contributions of different received features to the information recovery, further reflecting the distortion levels of the input features after transmission~\cite{cheng2023ml,cheng23c_interspeech}. \textcolor{black}{Specifically, as illustrated in~\cite{mi}, based on the chain rule and the smoothness of the estimated mutual information function, the correlation between the $i$-th output feature $\Bar{{d}}^j_i$ of neural network $\mathcal{F}$  and the loss $L_{\text{FDE}}$ can be derived as follows, 
\begin{equation}
\Delta L_{\text{FDE}}(\Bar{{d}}^j_i) \approx \nabla_{\Bar{{d}}^j_i} L_{\text{FDE}} \cdot \Delta \Bar{{d}}^j_i,    
\end{equation}
where $\nabla_{\Bar{{d}}^j_i} L_{\text{FDE}}$ represents the gradient of the loss function with respect to the output feature, $\Delta L_{\text{FDE}}(\Bar{{d}}^j_i)$ and $\Delta \Bar{{d}}^j_i$ are the variation of the loss function and the output feature after iterations, respectively. Additionally, the output feature variation $\Delta \Bar{{d}}^j_i$ is linearly related to its corresponding gradient, which can be represented as 
\begin{equation}
\label{feature variation-r}
\Delta \Bar{{d}}^j_i = -\eta \nabla_{\Bar{{d}}^j_i} L_{\text{FDE}},    
\end{equation}
where $\eta$ is the learning rate. Therefore, the variation of $L_{\text{FDE}}$ can be also represented as follows,
\begin{equation}
\Delta L_{\text{FDE}}(\Bar{{d}}^j_i) \approx -\eta \,\bigl\|\nabla_{\Bar{{d}}^j_i} L_{\text{FDE}}\bigr\|^2.    
\end{equation}
This suggests that if $\bigl|\nabla_{\Bar{{d}}^j_i} L_{\text{FDE}}\bigr|$ is larger, the loss decreases more rapidly, indicating that this feature contributes more to the reduction of the loss. As illustrated in~Eq.~(\ref{feature variation-r}), in the gradient descent process, the optimizer is more inclined to adjust these features, causing their variations to be more rapid to accelerate the reduction of the loss~\cite{adam}. Finally, the output feature vector of $\mathcal{F}$ can display a more pronounced distribution corresponding to the decrease in loss (\textcolor{black}{i.e., the increase in mutual information}), which can be used to guide the feature distortion evaluation.}

During testing, since the contributions of different received features to the information recovery are well-learned on all the categories, for an received feature vector $\hat{\bm{z}}_\mathrm{p}^j$ to feed into FDE, the output $\Bar{\bm{d}}^j$ can represent the semantic similarity of $\hat{\bm{z}}_\mathrm{p}^j$ relative to the undistorted vectors that the model have learned corresponding to the specific category. Consequently, the distortion levels of features in received feature vector $\hat{\bm{z}}_\mathrm{p}^j$ can be inferred from output vector $\Bar{\bm{d}}^j$ of FDE directly. Particularly, after applying the activation function to $\Bar{\bm{d}}^j$, larger values indicate greater similarities, implying less distortion of the corresponding features in $\hat{\bm{z}}_\mathrm{p}^j$. Finally, these values form the vector ${\widetilde{\bm{{d}}}}^j$ by selecting the corresponding positions according to $\hat{\bm{z}}^j$, and the distortion measure vector ${\bm{{d}}}^j$ can be given by ${\bm{{d}}}^j=1-{\widetilde{\bm{{d}}}}^j$.}  

Furthermore, to instruct the transmitter to retransmit the corrupted features, ${\bm{{d}}}^j$ is quantized into a binary feedback vector {${\bm{p}}^{(j+1)}$} to trigger the $(j+1)$-th retransmission. Note that the process is not trivial because the quantization is discrete, which makes the network non-differentiable and hard to optimize with back-propagation. Thus, a straight-through Gumbel-softmax estimator is investigated~\cite{gumbel-softmax}, where ${\bm{{d}}}^j$ is discretized in the forward pass but a gradient estimator based on Gumbel-softmax distributions is employed to resolve the non-differentiability in the backward pass. Specifically, for the forward propagation, {${\bm{p}}^{(j+1)}$} is drawn from ${\bm{{d}}}^j$ using a distortion threshold $t$, which can be represented as
{\begin{equation}\label{b}
{p}^{(j+1)}_i =\left\{\begin{array}{ll}
1 & \text { if } {d}^{j}_i \geq t, \\
0 & \text { otherwise, }
\end{array}\right.
\end{equation}}
where {${p}^{(j+1)}_i$} and {${d}^{j}_i$} denote the $i$-th value of {${\bm{p}}^{(j+1)}$} and {${\bm{{d}}^j}$}, respectively. On this basis, {${\bm{p}}^{(j+1)}$} can serve as a feedback vector to guide the next retransmission. Particularly, the number of features requested to be retransmitted is equal to the number of $1$ values in {${\bm{p}}^{(j+1)}$}, which is given by
{\begin{equation}\label{R}
R=\sum_{i=1}^B {p}^{(j+1)}_i,
\end{equation}}
{where $B$ is the size of {${\bm{p}}^{(j+1)}$}}. 

For the backward propagation, simply back-propagating through the binarization function as if it had been an identity function may cause discrepancies between the forward and backward pass, leading to high variance. Therefore, to improve the efficiency and stability of gradient estimation, Gumbel-softmax distributions are used, where each sample is a differentiable proxy of the corresponding discrete sample. Furthermore, note that straightforwardly applying single-dimension Gumbel-softmax is probable to sample repeatedly since the $R$ selections act independently and are not aware of each other. To solve this issue, a multi-dimension Gumbel-softmax scheme is proposed, where $R$ Gumbel-softmax distributions are designed simultaneously from $R$ categorical distributions generating from ${\bm{{d}}}^j$. The $R$ Gumbel-softmax distributions are used to estimate the gradients of $R$ samplings in the forward propagation. 

Specifically, a categorical vector ${\bm{{v}}}^{jr}\in\mathbb{R}^{B}$ is designed for the $r$-th sampling, where $r=1,...,R$. It is obtained from ${\bm{{d}}}^{jr}\in\mathbb{R}^{B}$ after a softmax function and can be denoted as
\begin{equation}
{\bm{{v}}}^{jr}= \mathrm{softmax}({\bm{{d}}}^{jr}),
\end{equation}
where ${\bm{{d}}}^{jr}$ is obtained after replacing the largest $r-1$ values with $0$ in ${\bm{{d}}}^j$. On this basis, ${\bm{{v}}}^{jr}$ serves as a probability vector involving the variables in ${\bm{{d}}}^{jr}$, and can be utilized to form the $r$-th Gumbel-softmax distribution. Since the samples from Gumbel-softmax distributions will become identical to the samples from the corresponding categorical distributions along with training, the maximum value in each categorical vector will most likely be sampled, matching the forward propagation. Therefore, ${\bm{{v}}}^{jr}$ is used to select the $r$-th largest value in ${\bm{{d}}}^j$, mitigating repetitive samplings. Consequently, the sample vector $\bm{w}^{jr}$ in $r$-th sampling is represented as
\begin{equation}
\label{sample vector}
\begin{split}
{\bm{{w}}}^{jr} = \mathrm{softmax}\biggl(\Bigl(\log({\bm{{v}}}^{jr})+\bm{g}^r\Bigl)/\tau\biggl),
\end{split}
\end{equation}
where $\bm{g}^r$ is a sampling vector drawn from Gumbel(0,1) distribution. The softmax function in Eq.~(\ref{sample vector}) is used as a continuous, differentiable approximation to $\arg\max$, and $\tau$ serves as the softmax temperature. As $\tau$ approaches $0$, the distribution will become more discrete and the sample vectors will converge to one-hot vectors. During training, the temperature parameter is decreased by an exponentially decreasing curve as in~\cite{multi-gs1,multi-gs2}. Finally, all the $R$ sample vectors are summed up to be a continuous relaxation of the quantized ${{\bm{{p}}}^{(j+1)}}$ for backward propagation, which is denoted as 
\begin{equation}
\bm w^{j}=\sum_{r=1}^R{\bm w}^{jr}.
\end{equation}

\subsection{Feature importance ranking~(FIR)}
\label{sec:importance} % 添加标签

\begin{figure}[tbp]
  \centering
  \centerline{\includegraphics[width=6.5cm]{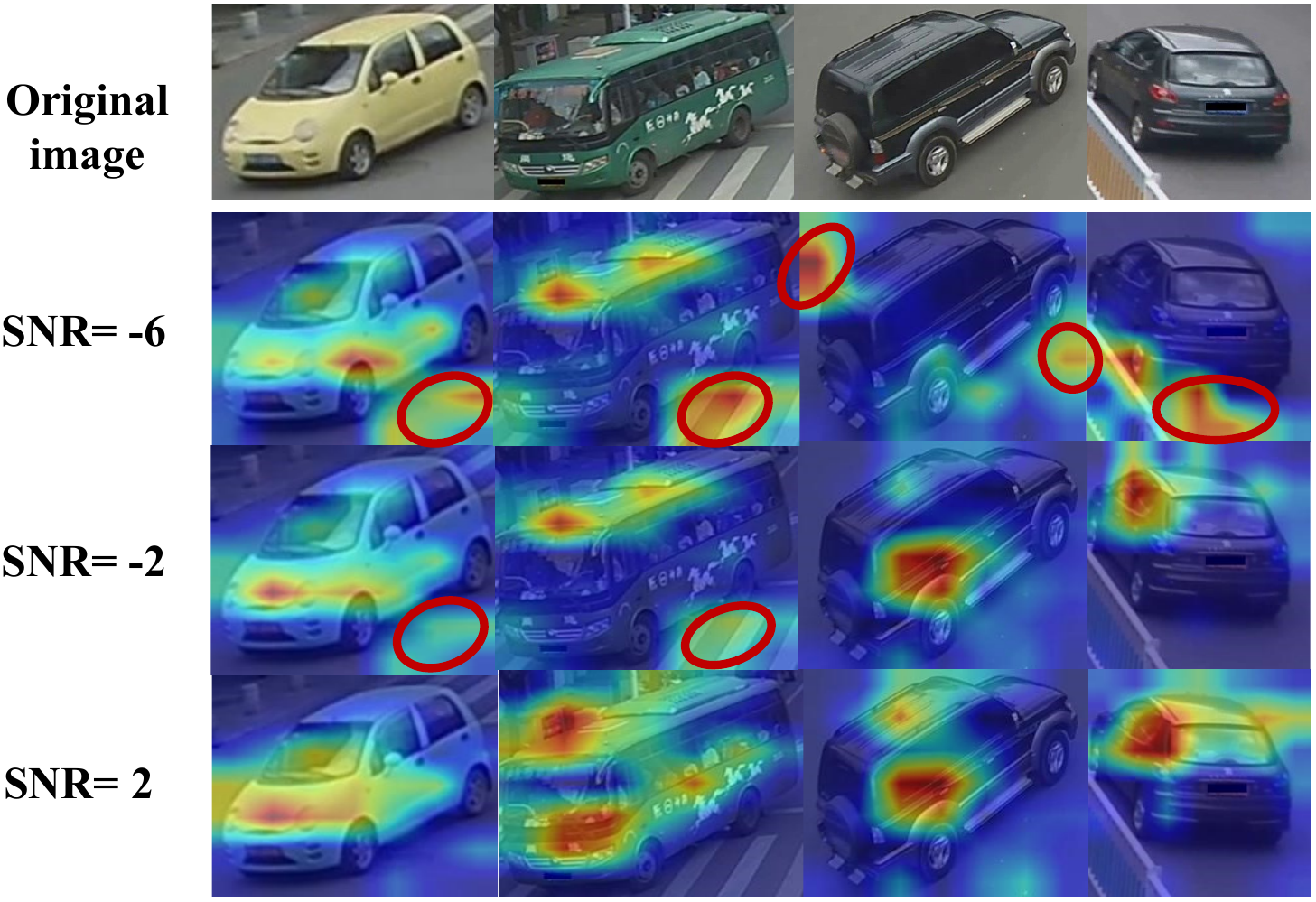}}
%  \vspace{2.0cm}
\caption{The gradient-weighted class activation mapping visualization results w.r.t. different channel conditions, where warmer colors indicate contributing semantic features while red circles indicate distracting backgrounds.}
\captionsetup[figure]{labelsep=space}
\label{heatmap}
\end{figure}

To identify the importance levels of different features in $\bm{f}$ w.r.t. specific tasks, we refer to gradient-weighted class activation mapping (Grad-CAM)~\cite{grad-cam1}, where visual explanations of the decision-making process of a CNN are proposed.  In Grad-CAM, the important regions within an input image are identified by leveraging the gradients flowing into the last convolutional layer of the CNN. The gradients represent the sensitivity of each feature map w.r.t. the target class. The higher the gradient value, the more influential the corresponding feature map is for the prediction. 

In this paper, the important features are identified by calculating the gradient of the task predictions relative to the feature vector $\bm{f}$. Intuitively, Fig.~{\ref{heatmap}} demonstrates the visualization results with the proposed SemHARQ architecture of three multi-task, \textcolor{black}{i.e., vehicle re-identification (ReID)}, vehicle color classification and vehicle type classification, where warmer colors indicate regions with more significance. Apparently, as SNR gets better, SemHARQ focuses more on the vehicles than the backgrounds, leading to better task performance. The specific methodology for calculating the importance levels of semantic features is shown as follows.

\begin{figure}[tbp]
  \centering
  \centerline{\includegraphics[width=7cm]{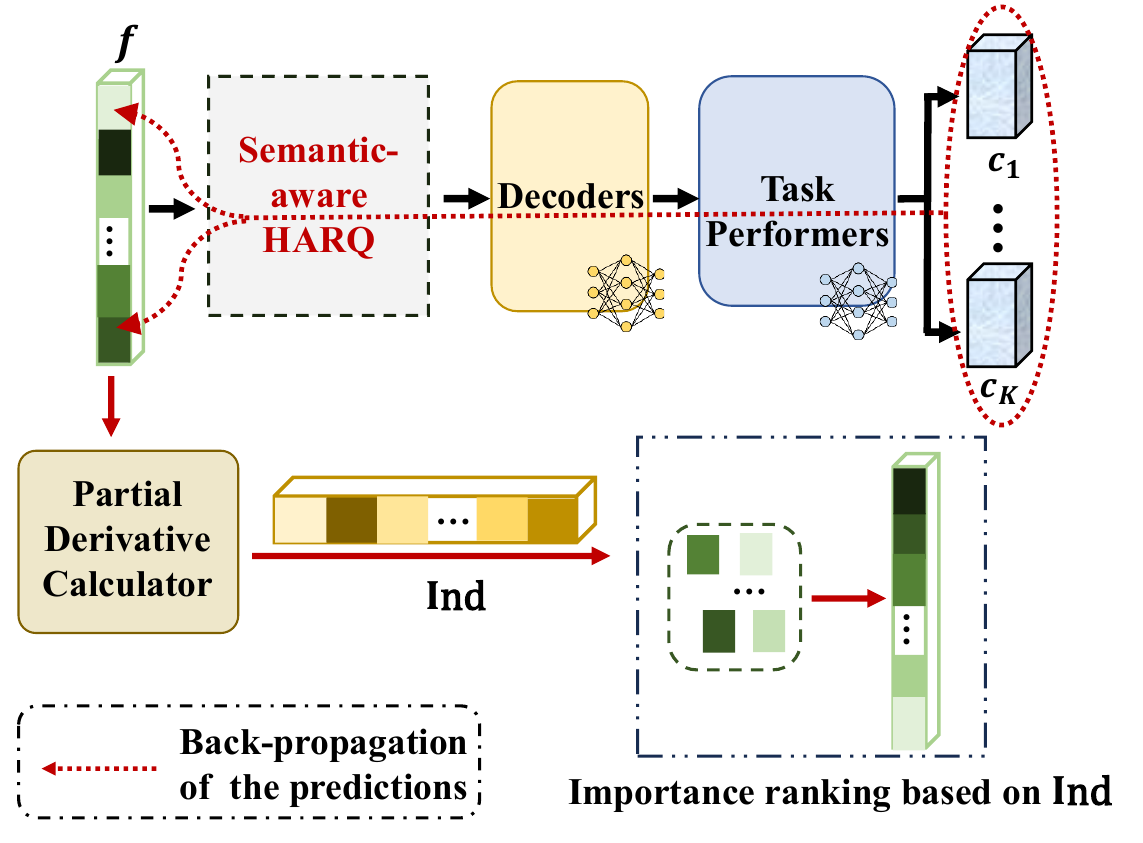}}
%  \vspace{2.0cm}
\caption{
{The framework of FIR, where features with deeper colors in $\bm{f}$ attach higher importance.}}
\captionsetup[figure]{labelsep=space}
\label{FIR}
\end{figure}

Firstly, as shown in Fig.~\ref{FIR}, the input images are fed into the proposed SemHARQ to obtain the output logits w.r.t. different tasks, where the entire $\bm{f}$ is encoded and transmitted without feature selection. Retransmissions are triggered and guided by the FDE network if the received features fail to pass the detection of the retransmission identification module. Then, the gradients of the target classes w.r.t. the features of $\bm{f}$ are calculated by back-propagating the final predictions through the system. Specifically, the probability vector output by the $k$-th task performer is denoted as $\bm{c}_k\in\mathbb{R}^M$, where $M$ indicates the total number of categories of the $k$-th task. The importance level corresponding to each feature in $\bm{f}$ is obtained by calculating the partial derivative of ${c}_{km}$ relative to $\bm{f}$, where ${c}_{km}$ is the probability corresponding to the predicted category. The importance measure vector w.r.t $k$-th task is then represented as
\begin{equation}
\label{deqn_ex1}
{\bm{\widetilde{f}}_k}=\frac{\partial {{c}_{km}} }{ \partial{\bm{f}}},
\end{equation}
which represents the significance of each feature of $\bm{f}$ to the $k$-th task. Finally, the importance measure vectors of all the $K$ tasks are summed up, and then an integral importance measure vector is obtained,
\begin{equation}
\label{deqn_ex1}
{\textbf {Ind}}=\sum_{k=1}^K\lambda_k\bm{\widetilde{f}}_k,
\end{equation}
where $\lambda_k$ denotes the weight of the $k$-th task, $\bm{\widetilde{f}}_k$ is the importance measure vector corresponding to the $k$-th task. In the end, the features in $\bm{f}$ are selected based on the importance measure vector ${\textbf {Ind}}$ under various channel conditions. The features corresponding to the positions with larger values in ${\textbf {Ind}}$ are with higher priority at the first transmission and all the incremental transmissions.

\section{Semantic-aware HARQ}
In this section, the process of SemHARQ is further analyzed. {Note that to transmit the meanings of the source efficiently, all the functions of SemHARQ are performed at the level of semantic features. The JSCC is utilized in our framework to encode semantic features directly into channel transmission symbols.} Furthermore, the training strategy of SemHARQ is proposed to optimize the training process.

\subsection{SemHARQ}

\label{sem-harq}
\begin{figure}[tbp]
\centering
 \centerline{\includegraphics[width=8cm]{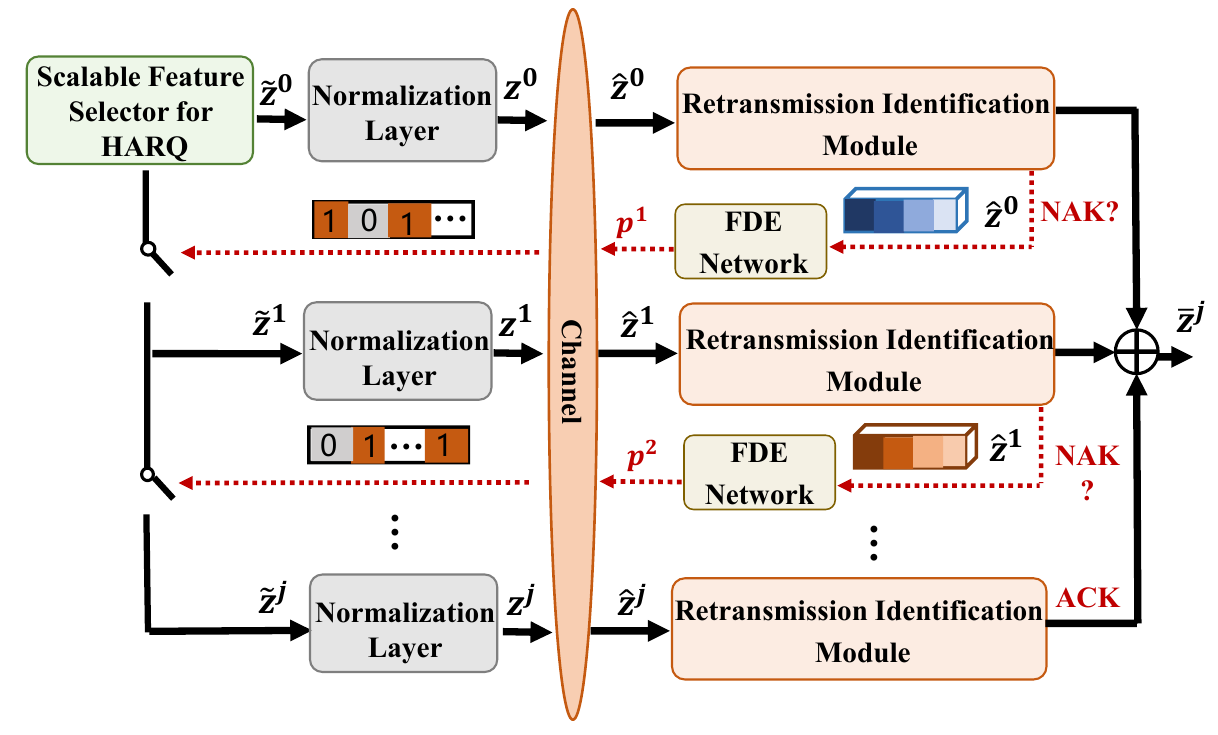}}
\caption{The working flow of SemHARQ.}
\captionsetup[figure]{labelsep=space}
\label{HARQ}
\end{figure}

\begin{figure}[t]
\centering
\centerline{\includegraphics[width=0.3\textwidth]{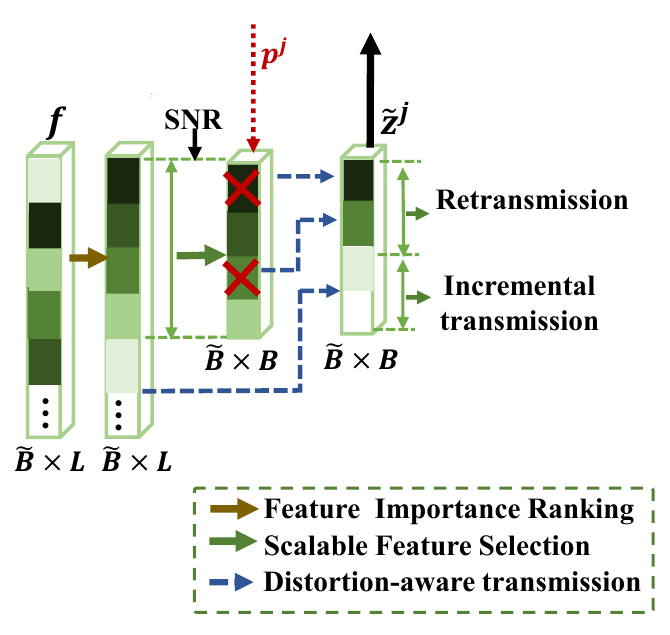}}
%\captionsetup{justification=centering}
\caption{The structure of the scalable feature selector for SemHARQ.}
%\captionsetup[figure]{labelsep=space}
\label{sfs}
\end{figure}

The structure of SemHARQ of $j$ retransmissions is shown in Fig.~\ref{HARQ}. Firstly, the feature vector to be transmitted at the first transmission is obtained by encoding and scalable feature selection, which is represented by
\begin{equation}
\bm{\widetilde{z}}^0 = {f}_{\text{sfs}}\biggl(\emph{C}_{\alpha}\Bigl(\emph{S}_{\beta}(\bm{x})\Bigl)\biggl),
\end{equation}
where $\emph{S}_{\beta}(\cdot)$ and $\emph{C}_{\alpha}(\cdot)$ denote the multi-task semantic encoder with the parameter set $\beta$ and the JSC-Encoder with the parameter set $\alpha$, respectively, and $f_{\text{sfs}}(\cdot)$ indicates the scalable feature selector, whose structure is shown in Fig.~\ref{sfs}. 
Consequently, $\bm{\widetilde{z}}^0$ consists of $B$ features chosen from the encoded feature vector according to the importance measure vector ${\textbf {Ind}}$ and the given channel condition. After normalizing and passing through the channel, the received feature vector is represented as $\hat{\bm{z}}^0$ at the receiver. 

Then, $\hat{\bm{z}}^0$ is detected by the retransmission identification module to determine the need for retransmission. Specifically, a retransmission criterion based on the received data uncertainty is applied~\cite{retransmission1}. Consider the acquisition of a concatenated feature vector $\overline{\bm{z}}^j$ after $j$ retransmissions, the receiver repeatedly requests the transmitter to retransmit until
\begin{equation}\label{criteria_equa}
\operatorname{\overline{SNR}}>\min \biggl (\theta_{0}\mathcal{L}\Bigl(\mathcal{U}({\bm{\overline{z}}^j})\Bigl) , \theta_{\mathrm{SNR}}\biggl),
\end{equation}
where $\operatorname{\overline{SNR}}$ stands for the receive-$\text{SNR}$ after the previous transmissions, $\mathcal{U}(\cdot)$ denotes the received information uncertainty, $\mathcal{L}(\cdot)$ represents a monotonically increasing reshaping function which is set to $\mathcal{L}(x)=1+x$, $\theta_{0}$ indicates a given conversion ratio between the uncertainty and $\text{SNR}$ of the combined information, and $\theta_{\mathrm{SNR}}$ is a given maximum receive-$\text{SNR}$. \textcolor{black}{Particularly, the receive-$\text{SNR}$ is obtained by integrating the information after multiple retransmissions, which is given by 
\begin{equation}\label{receive SNR form}
\operatorname{\overline{SNR}}=\frac{2 P}{\sigma^{2}}\sum_{j=0}^{J}\left\|\bm{h}^{j}\right\|^{2},
\end{equation}
where $P$ is the average power of the semantics, $\bm{h}^{j}$ represents the channel coefficient vector {at the $j^{th}$ retransmission}}, $\frac{\sigma^{2}}{2}$ indicates the noise variance, $J$ denotes the total number of retransmissions. As for the uncertainty, $\mathcal{U}({\bm{\overline{z}}^j})$ is measured by the average entropy of the posterior probability distribution corresponding to different tasks. In terms of the $k$-th task, the uncertainty is given by
\begin{equation}\label{uncertainty}
\mathcal{U}_{k}({\bm{\overline{z}}^j})=-\sum_{y_{k}} p_{\gamma}(y_{k} \mid {\bm{\overline{z}}^j}) \log p_{\gamma}(y_{k} \mid {\bm{\overline{z}}^j}),
\end{equation}
where $y_{k}$ denotes a class label and $\gamma$ is the set of parameters of the decoders and the task performer to be learned. The uncertainty measures the contribution of the received semantics to the task performance from the perspective of information entropy, and thus serves as a criterion for measuring the adequacy of the amount of received information. Furthermore, since the uncertainty  will be constantly bridged by increasing the receive-$\text{SNR}$  or equivalently the number of retransmissions, the inequality in Eq.~(\ref{criteria_equa}) is intuitively used to evaluate the adequacy of the number of retransmissions.

After passing through the retransmission identification module, if  $\hat{\bm{z}}^0$ passes the detection successfully, it is fed into the decoders and the multiple task performers to generate the multi-task predictions. The received feature vector is decoded as 
\begin{equation}
{\bm{c}_k} = \emph{T}_{\phi_{k}}\biggl(\emph{S}^{-1}_{\beta_{k}^{-1}}\Bigl(\emph{C}^{-1}_{\alpha^{-1}}({\hat{\bm{z}}^0})\Bigl)\biggl),
\end{equation}
where $\bm{c}_k$ denotes the predicted probability vector of the $k$-th task, $\emph{C}^{-1}_{\alpha^{-1}}(\cdot)$, $\emph{S}^{-1}_{\beta_{k}^{-1}}(\cdot)$ and $\emph{T}_{\phi_{k}}(\cdot)$  represent the JSC-Decoder with parameter set ${\alpha^{-1}}$ , the semantic decoder of the $k$-th task with parameter set ${\beta_{k}^{-1}}$ and the task performer of the $k$-th task with parameter set  ${\phi_{k}}$, respectively. {Additionally, if $\hat{\bm{z}}^0$ fails to pass the detection, the FDE network is utilized to generate the feedback vector $\bm{p}^1$ to trigger the first distortion-aware retransmission. The distorted features in $\hat{\bm{z}}^0$ are required to be retransmitted according to the distortion evaluation. Note that the number of transmitted features at every transmission is adaptive to the channel conditions. Therefore, the incremental transmission of new features in descending order of importance levels is combined with the retransmission if there are remaining channel resources. On this basis, the transmission robustness and efficiency are enhanced simultaneously.} Furthermore,  after the first retransmission, $\hat{\bm{z}}^1$ is obtained at the receiver and concatenated with the cached semantics from the previous transmission. The retransmission process will continue until either the receive-$\text{SNR}$ is sufficiently large or the predetermined maximum number of retransmissions is reached.

\begin{algorithm}[t]
	\caption{Training process of SemHARQ.} 

	\label{alg}
        
        \textbf{Stage 1:} Training of the FDE network.
    \begin{algorithmic}[1]
        \Require The semantic encoded $\bm{s}$ and JSC-decoded $\bm{\hat{s}}$, initialized parameters $\xi_{\mathcal{F}}$ of the distortion measure module $\mathcal{F}$ in the FDE network.
        \Ensure  $\xi_{\mathcal{F}}$.
        \State  Sample from $J(\bm{s},\bm{\hat{s}})$ and $M(\bm{s},\bm{\hat{s}})$, and feed the samples into $\mathcal{F}$. 
        \State  Estimate the mutual information of $\bm{s}$ and $\bm{\hat{s}}$ by Eq.~(\ref{estimated mi}).
         \State  Train $\xi_{\mathcal{F}}$ by loss function $-I(\bm{s};\bm{\hat{s}}|\mathcal{F})$.
    \end{algorithmic}

    \textbf{Stage 2:} Training of the whole HARQ system without scalable feature selection.
    \begin{algorithmic}[1]
        \Require Images $\bm{x}$ from training dataset, initialized parameters $\alpha$, $\beta$, $\alpha^{-1}$, $\beta_k^{-1}$, $\phi_k$ with $\bm{f}$ fully transmitted; $k = 1,...,K$.
        \Ensure  $\emph{S}_{\beta}(\cdot)$, $\emph{C}_{\alpha}(\cdot)$, $\emph{C}_{\alpha^{-1}}^{-1}(\cdot)$, $\emph{S}^{-1}_{\beta_{k}^{-1}}(\cdot)$, $\emph{T}_{\phi_{k}}(\cdot)$.
        \State  \textbf{Transmitter:} $\bm{f}= \emph{C}_{\alpha}\Bigl(\emph{S}_{\beta}(\bm{x})\Bigl)$.
                \While {the retransmission criterion and the maximum number of retransmissions are not met}
                \State Obtain $\bm{p}^j$ from the FDE network and 
                \State Send $\bm{p}^j$ back to the transmitter.
                \State Trigger distortion-aware retransmission according to $\bm{p}^j$.
                \EndWhile
        \State  \textbf{Receiver:}  ${\bm{\overline{z}}}^j=[{\bm{\hat{z}}}^1,...,{\bm{\hat{z}}}^j]$, ${\bm{c}_k} = \emph{T}_{\phi_{k}}\Bigl(\emph{S}^{-1}_{\beta_{k}^{-1}}(\emph{C}^{-1}_{\alpha^{-1}}(\bm{\overline{z}}^j))\Bigl)$.
        \State  Train $\alpha$, $\beta$, $\alpha^{-1}$, $\beta_k^{-1}$, $\phi_k$ by loss function $\emph{L}_{\text{E2E}}$ ; $k = 1,...,K$.
        \State Calculate the importance measure vector ${\textbf {Ind}}$.
    \end{algorithmic}

    \textbf{Stage 3:} Retraining of the SemHARQ with scalable feature selection according to ${\textbf {Ind}}$.
     
\end{algorithm}

\subsection{Training strategy}
\label{training strategy}
To improve the efficiency of SemHARQ training, a three-stage training strategy is proposed. We focus on three representative tasks in intelligent traffic scenarios for experiments, namely vehicle ReID, vehicle color classification, and vehicle type classification. \textcolor{black}{DenseNet-121 pre-trained on \texttt{ImageNet} is used as the initial model~{\cite{dense}}}. In the first stage, to improve the performance of feature distortion evaluation, the FDE network is trained with other trainable modules frozen. 
The training loss can be expressed as, 
\begin{equation}
\label{Lfde}
L_{\text{FDE}}=L_{\text{CH}}+L_{\text{MI}},
\end{equation}
where $L_{\text{CH}}$, $L_{\text{MI}}$ denote the channel transmission loss and the loss of mutual information, respectively. Specifically, $L_{\text{CH}}$ can be measured by the mean square error (MSE),
\begin{equation}
\label{deqn_ex1}
L_{\text{CH}}=\frac{1}{N}\sum_{n=1}^N||{s}_n-{\hat{s}}_n||^2_2,
\end{equation}
where $\bm{s}$ denotes the feature vector extracted by the semantic encoder, and $\bm{\hat{s}}$ is the feature vector decoded by the JSC-Decoder. The mutual information loss is denoted as
\begin{equation}
\label{deqn_ex1}
L_{\text{MI}}=-I(\bm{s};\bm{\hat{s}}|\mathcal{F}),
\end{equation}
where $I(\bm{s};\bm{\hat{s}}|\mathcal{F})$ is calculated in Eq.~(\ref{estimated mi}).

Then, in the second stage, to calculate the feature importance levels, the whole system is trained with the entire $\bm{f}$ transmitted without considering the constraint of communication capacity. The overall E2E loss is represented by
\begin{equation}
\label{deqn_ex1}
L_{\text{E2E}}=L_{\text{FDE}}+L_{\text{T}},
\end{equation}
where $L_{\text{T}}$ is the multi-task loss. Specifically, the losses of the three tasks are weighted, which can be given by
\begin{equation}
\label{deqn_ex1}
L_{\text{T}}=\lambda_{r}L_{r}+\lambda_{c}L_{c}+\lambda_{t}L_{t},
\end{equation}
where $L_{r}$, $L_{c}$, and $L_{t}$ correspond to the ReID, color classification, and type classification losses, respectively, $\lambda_{r}$, $\lambda_{c}$, and $\lambda_{t}$ denote the respective task weights. Specifically, $L_{r}$ consists of the hard-mining triplet loss~\cite{tri-loss} and the cross-entropy loss, defined as $L_{r}=\mathcal{L}_{ht}(a,p,n)+\mathcal{L}_{ce}(\bm{y},\bm{\hat{y}})$, where $a,p$ and $n$ represent anchor, positive and negative elements, respectively. The hard-mining triplet loss $\mathcal{L}_{ht}(a,p,n)$ is defined as
\begin{equation}
\label{deqn_ex1}
\mathcal{L}_{ht}(a,p,n)= \max\left\{\Bigl(\max(d_{ap})-\min(d_{an})+\alpha\Bigl),0\right\},
\end{equation}
where $d_{ap}$ and $d_{an}$ denote the distances between the extracted features of the anchor and positive/negative images. $\mathcal{L}_{ce}(\bm{y},\bm{\hat{y}})$ indicates the cross-entropy loss, which is defined as
\begin{equation}
\label{deqn_ex1}
\mathcal{L}_{ce}(\bm{y},\bm{\hat{y}})=\sum_{k=1}^K{y}_k\log({\hat{{y}}}_k)+(1-{{y}}_k)\log(1-{\hat{{y}}}_k),
\end{equation}
where $\bm{y}$ denotes the label of the task and $\bm{\hat{y}}$ is the predicted one. For the other two classification tasks, the cross-entropy loss is also employed.

\newcommand{\tabincell}[2]{\begin{tabular}{@{}#1@{}}#2\end{tabular}}  

\begin{table*}[tbp]
  \renewcommand{\arraystretch}{1.2} % 可以调节, 1.2指高度是默认的1.2倍
  \centering
  
  \caption{The network settings of SemHARQ.}
\setlength{\tabcolsep}{4.5mm}{
\begin{threeparttable}
\begin{tabular}{c|c|c|c|c}
\hline
{Transmission System} & {Modules} & {Layer} & \tabincell{c}{Output\\dimensions} & \tabincell{c}{Activation\\function}\\ \hline
\multirow{7}{*} {Transmitter} & Input  & $\bm{x}$  & {$256\times 256$}  & / \\  \cline{2-5} 
 & {Multi-task semantic encoder}  & \tabincell{c}{DenseNet-121}  & \tabincell{c}{1024} & \tabincell{c}{ReLu} \\ \cline{2-5} 
 & JSC-Encoder & \tabincell{c}{FC with BN\\ FC} &  \tabincell{c}{1024\\1024} &\tabincell{c}{Leaky ReLu\\Linear} \\ 
 \cline{2-5} 
 & Scalable feature selector for HARQ & \tabincell{c}{/} &  \tabincell{c}{$B$} &\tabincell{c}{/} \\ 
 \cline{2-5}
 & \tabincell{c}{Output ($j^{th}$ retransmission)}  & $\bm{\widetilde{z}}^j$  & $B$ & / \\ 
 \cline{2-5} \hline 
  \rowcolor{gray!20} 
Channel & Rician fading & /& /& / \\\hline 

\multirow{11}{*}{\tabincell{c}{Receiver}} 
& Input  & $\bm{\hat{z}}^j$  & $B$ & / \\ \cline{2-5}
& Retransmission identification  & NACK or concatenation  & / or ${B\times(j+1)}$ & / \\ \cline{2-5} 
 & JSC-Decoder & \tabincell{c}{FC with BN\\ FC} &  \tabincell{c}{1024\\1024} &\tabincell{c}{Leaky ReLu\\ReLu} \\  \cline{2-5} 
 & Semantic decoder $1$ for vehicle ReID & \tabincell{c}{FC} & \tabincell{c}{1024} &\tabincell{c}{Leaky ReLu}  \\ \cline{2-5}  
 & Semantic decoder $2$ for vehicle color classi. & \tabincell{c}{FC} & \tabincell{c}{512} &\tabincell{c}{Leaky ReLu}  \\ \cline{2-5} 
  & Semantic decoder $3$ for vehicle type classi. & \tabincell{c}{FC} & \tabincell{c}{512} &\tabincell{c}{Leaky ReLu}  \\ \cline{2-5}
& Task performer $1$ for vehicle ReID & \tabincell{c}{FC} & \tabincell{c}{575} &\tabincell{c}{/}  \\ \cline{2-5}
 & Task performer $2$ for vehicle color classi. & \tabincell{c}{FC} & \tabincell{c}{10} &\tabincell{c}{/}  \\ \cline{2-5}
 & Task performer $3$ for vehicle type classi. & \tabincell{c}{FC} & \tabincell{c}{9} &\tabincell{c}{/}  \\ \cline{2-5}
 & Output  & {$\bm{c}_1$}, {$\bm{c}_2$}, {$\bm{c}_3$}   & 575, 10, 9 & / \\  \hline

\multirow{4}{*}{\tabincell{c}{FDE network\\ ($j^{th}$ retransmission)}}
&  $\mathcal{F}$  & \tabincell{c} {Dense Layer 1\\ Dense Layer 2\\ Dense Layer 3\\ Dense Layer 4}  & \tabincell{c} {1024\\256\\1024\\1} & \tabincell{c}{ReLu\\ReLu\\Sigmoid\\ / }\\ \cline{2-5} 
  &  Mutual information estimation  & \tabincell{c} {/}  & \tabincell{c} {1} & \tabincell{c}{ / }\\ \cline{2-5} 
  &  Multi-dimension Gumbel-softmax  & \tabincell{c} {$\bm{p}^j$}  & \tabincell{c} {$B$} & \tabincell{c}{/ }\\  
  \hline

\end{tabular}
\begin{tablenotes}
\item[1] FC: fully connected layer; BN: batch normalization; ReLu: rectified linear unit.
\item[2] The retransmission identification module sends NACK to the FDE network if the detection is not successful, otherwise, concatenates the received vector with the buffered feature vectors.
\end{tablenotes}
\end{threeparttable}}
\end{table*}

Finally, the SemHARQ considering scalable feature selection is trained based on the importance measure vector ${\textbf {Ind}}$ with limited bandwidth and time slot, where all the trainable parameters are optimized. The most important $B$ features are transmitted, where $B$ is determined by channel conditions. The pseudocode of the whole training process is shown in Algorithm~\ref{alg}.

\section{Numerical Results}
In this section, simulations are performed to evaluate the proposed system. The details of the simulation settings and the structure of the network are introduced. Then, the proposed SemHARQ is compared with other baselines, and the results are analyzed.

\subsection{Dataset and performance metrics}
The adopted dataset is VeRi-$776$~\cite{liu2016deep}, which contains over $50,000$ images of $776$ vehicles captured by $20$ cameras covering an area of $1.0$ square kilometers, making the dataset flexible to be used for vehicle ReID and other related tasks. The images are captured in unconstrained surveillance scenarios and are labeled with different attributes such as vehicle type, color, and brand. Thus, the evaluation results are all presented based on the VeRi-$776$ dataset.

Two of the most popular metrics, rank-1 accuracy and mean average precision (mAP), are used to evaluate the performance of vehicle ReID. Specifically, rank-1 accuracy measures the percentage of query images that are correctly matched at the top-ranked position based on the similarity or distance metrics. The mAP represents the mean of the AP, indicating the proportion of the correctly retrieved gallery images in the gallery image set. It evaluates the identification performance in a more global view. Additionally, the other two classification tasks are evaluated by the average accuracy.

\subsection{Simulation settings and numerical analysis}
\begin{figure}[t]
\centering
\centerline{\includegraphics[width=0.45\textwidth]{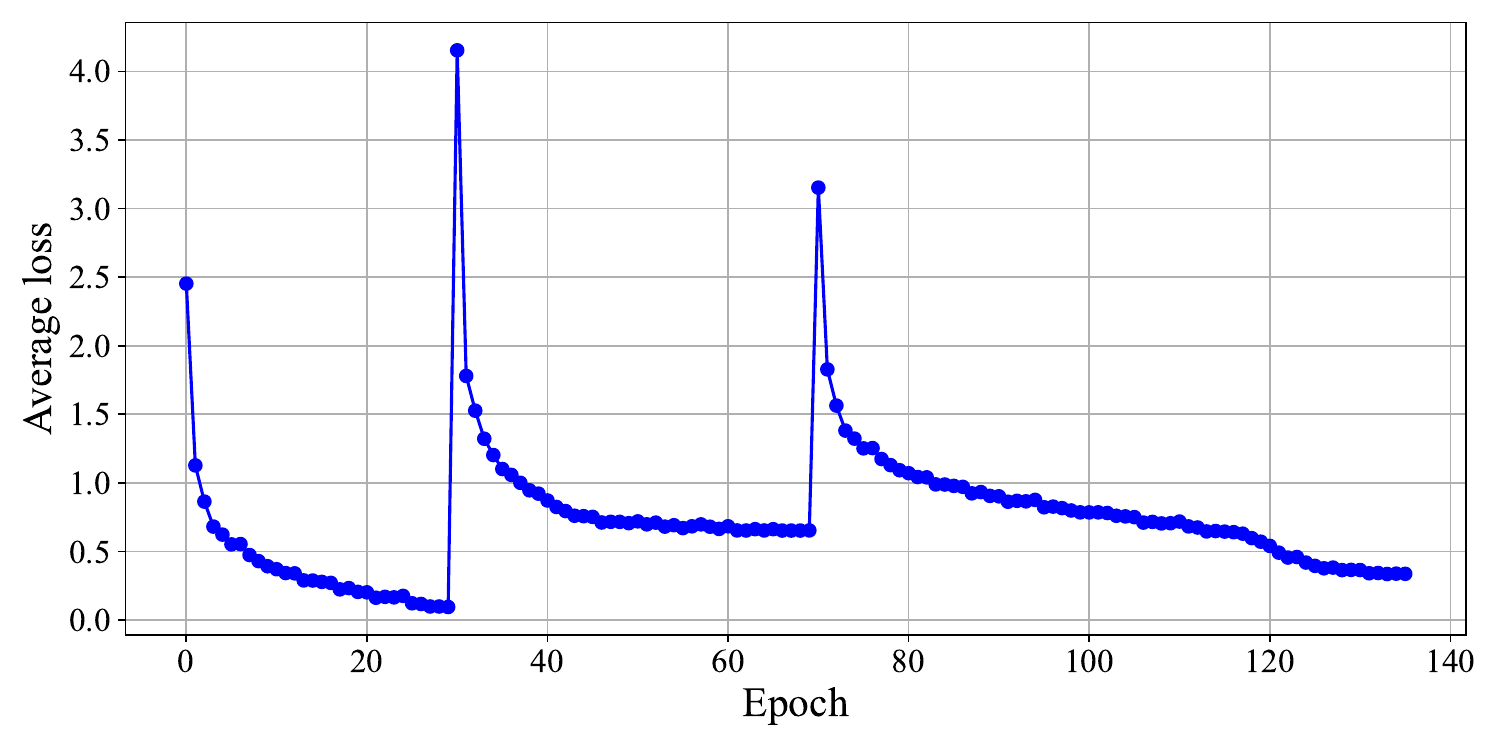}}
%\captionsetup{justification=centering}
\caption{The loss curve of SemHARQ training with SNR = 0~dB.
}
%\captionsetup[figure]{labelsep=space}
\label{loss}
\end{figure}

The detailed network settings of the proposed SemHARQ are shown in Table I. The system is evaluated under several SNR levels and the average power budget P is set to $1$. DenseNet-121 is utilized as our backbone deep neural network~(DNN), which is optimized by {\texttt{Adam}}~\cite{adam}. $\lambda_{r}$, $\lambda_{c}$, $\lambda_{t}$ are set to $1, 0.125, 0.125$, respectively. \textcolor{black}{For a given SNR, the entire training process consists of 135 epochs, including 30 epochs for pre-training the FDE network to estimate feature distortion, followed by 40 epochs for training the whole SemHARQ system without scalable feature selection to obtain feature importance scores, and another 65 epochs for retraining the SemHARQ model with scalable feature selection based on the learned importance scores. The training batch and testing batch are set to 32 and 100, respectively. Both training and testing are conducted on a single NVIDIA V100 GPU. The entire model is trained using a dataset of approximately 40,000 RGB images with a resolution of $3\times 256 \times 256$. \textcolor{black}{To further illustrate the convergence behavior of our model, a training loss curve of SemHARQ training with SNR = 0~dB is shown in Fig.~\ref{loss}. It can be observed that the training loss decreases smoothly and steadily during the three training phases as discussed in Algorithm~\ref{alg}, eventually converging around epoch 25, epoch 60 and epoch 130, respectively. } Meanwhile, around 3.6 GB of memory is consumed during testing on images with a resolution of $3\times 256 \times 256$. Note that techniques such as model pruning~\cite{fang2023depgraph} and knowledge distillation~\cite{wei2024scaled} can be adopted to further reduce memory consumption in practical applications.}  Besides, the learning rate is initialized as 1e$-4$, and learning rate decay is applied. In addition, the conversion ratio between the uncertainty and the receive-SNR of the combined information $\theta_0$ in Eq.~(\ref{criteria_equa}) is set to $1$ dB, and the maximum receive-SNR $\theta_{\text{SNR}}$ can be determined by $\theta_{\mathrm{SNR}}=\theta_{0}\left[1+ \log{M}\right]$, where $\log{M}$ is the maximum entropy, with $M$ denoting the total number of categories. {Moreover, to trade off the robustness and the diversity of the received semantics, the number of features to be retransmitted is approximately the same as those for incremental transmission on average by adjusting the distortion threshold $t$ in Eq.~(\ref{b}). Note that the value of $t$ can also be readily changed. Particularly, to explore the impact of $t$, the multi-task performance of different retransmission ratios after adjusting $t$ is shown in Table~\ref{t ratio}, where the retransmission ratio represents the average ratio of the retransmitted features to all the received features. It is obvious that better performance is achieved when the ratio is set to $0.5$. The reason is that the semantic error correction and new information transmission are better traded off, enhancing the communication efficiency.} 

\begin{table}[t]
\caption{{Multi-task accuracy of different retransmission ratios.}}
\large
\scriptsize
\centering
\tabcolsep=0.1cm
\renewcommand\arraystretch{1.4}
{
\begin{tabular}{lccccccccc}
\toprule
 & \multicolumn{9}{c}{Retransmission ratio} \\
\cmidrule(lr){2-10}
{SNR~[dB]}& \multicolumn{3}{c}{0.25} & \multicolumn{3}{c}{0.5} & \multicolumn{3}{c}{0.75} \\
\cmidrule(lr){2-4}\cmidrule(lr){5-7}\cmidrule(lr){8-10}
& Rank-1 & Color & Type & Rank-1 & Color & Type & Rank-1 & Color & Type \\
\midrule
$-6$ & 17.61 & 88.09 & 76.48 & 20.97 & 88.95 & 80.68 & 18.85 & 88.10 & 78.86 \\
$-4$ & 19.45 & 89.69 & 84.30 & 23.65 & 90.81 & 85.15 & 20.89 & 89.75 & 85.26 \\
$-2$ & 60.98 & 90.49 & 88.47 & 68.22 & 91.30 & 88.81 & 63.35 & 90.75 & 88.55 \\
$0$  & 72.94 & 91.17 & 90.07 & 76.94 & 92.17 & 91.07 & 73.69 & 92.02 & 90.84 \\
$2$  & 85.11 & 93.19 & 91.79 & 85.74 & 93.75 & 91.64 & 85.00 & 92.25 & 91.53 \\
$4$  & 89.78 & 93.51 & 91.92 & 89.94 & 93.78 & 91.75 & 89.90 & 93.06 & 91.80 \\
$6$  & 90.40 & 93.61 & 92.36 & 90.58 & 94.21 & 91.85 & 90.41 & 93.40 & 92.32 \\
$8$  & 90.41 & 93.72 & 92.23 & 91.54 & 94.15 & 92.52 & 91.03 & 93.80 & 92.55 \\
\bottomrule
\end{tabular}}
\label{t ratio}
\end{table}

The proposed E2E system is compared with the state-of-the-art HARQ benchmarks. The details are as follows:

\makeatletter
\newcommand{\rmnum}[1]{\romannumeral #1}
\newcommand{\Rmnum}[1]{\expandafter\@slowromancap\romannumeral #1@}
\makeatother

\begin{itemize}
		\item[$\bullet$] {Traditional $\text{\Rmnum{2}}$-HARQ method: JPEG, LDPC with $\frac{3}{4}$ rate and $16$ QAM are used for source coding, channel coding and modulation, respectively. Then, the CRC is employed at the receiver and the incremental redundancy information is requested to transmit when CRC fails until the success or the maximum number of transmissions is reached. After the image reconstruction, the three intelligent tasks are performed using DenseNet121.}
		\item[$\bullet$] {Semantic-aware HARQ baselines: 1) IK-S-HARQ~\cite{IB-harq}: An incremental knowledge-based semantic-aware HARQ method is compared. Specifically, the $B$ features to be transmitted are incrementally selected from $\bm{f}$ at each transmission in order of the importance level. 2) RT-S-HARQ~\cite{retransmission1}: A retransmission-only semantic-aware HARQ method is compared, where the most important $B$ features are selected and retransmitted at each transmission.
                          }
\end{itemize}
{Note that the comparison only focuses on the HARQ method at the semantic feature level, which is independent of the source modality.  The proposed SemHARQ and the semantic-aware HARQ baselines are all DL-based with the same backbone DNN, \textcolor{black}{i.e., DenseNet-121}. Meanwhile, the JSCC and scalable feature selector with the same structures are applied to achieve the best results. }

\begin{figure*}[tbp]
	\centering
	\subfloat[{Rank-1 accuracy of different methods}]{\label{fig:a}\includegraphics[width=6cm]{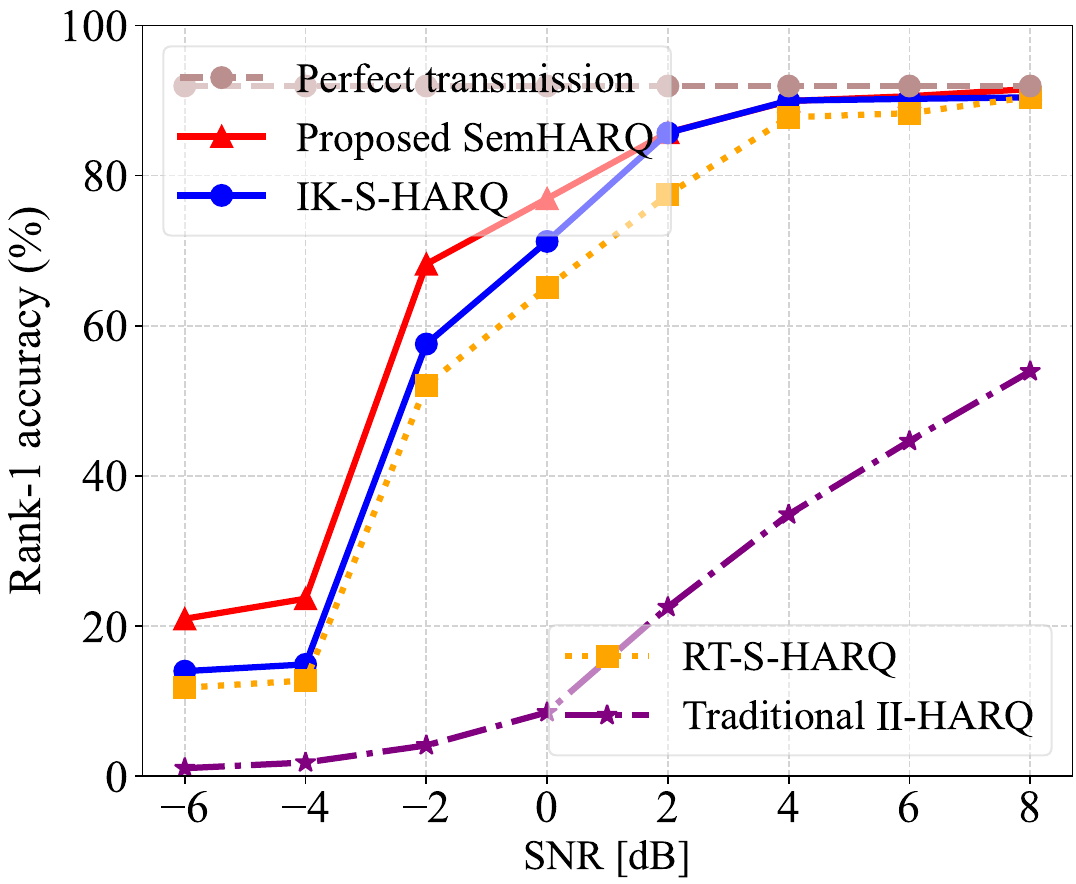}}\label{Fig_rank}\quad
   \vspace{1mm}
	\subfloat[{mAP of different methods}]{\label{fig:b}\includegraphics[width=6cm]{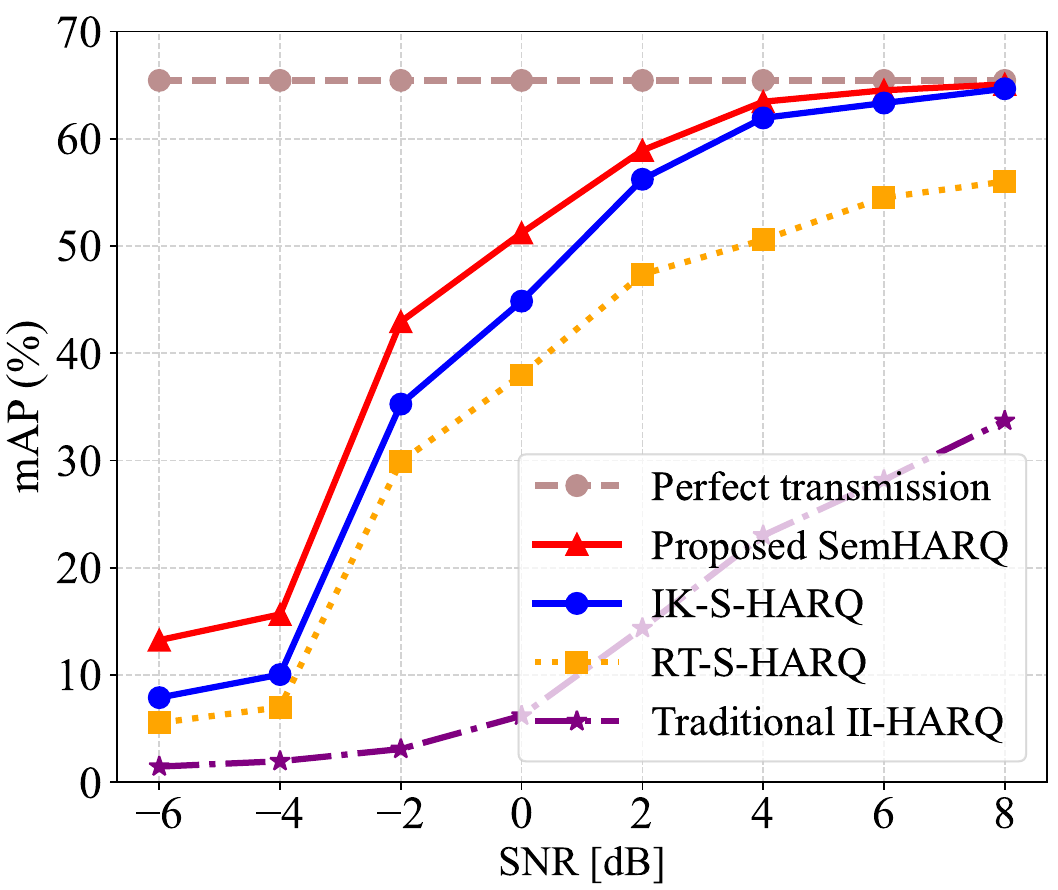}}\label{Fig_map}\\	
	\caption{ {Rank-1 accuracy and mAP of vehicle ReID w.r.t. different methods. }}
	\label{Fig_vehicle_reid}
\end{figure*}

Fig.~\ref{Fig_vehicle_reid} presents the vehicle ReID performance w.r.t. different methods. It is obvious that the semantic-aware HARQ methods outperform the traditional method in both rank-1 accuracy and mAP, which verifies the superiority of SemCom in transmitting the meaning of information efficiently. Furthermore, the proposed SemHARQ achieves the best performance among the three semantic-aware HARQ methods, followed by IK-S-HARQ and RT-S-HARQ. {The rank-1 accuracy gaps between SemHARQ and IK-S-HARQ and RT-S-HARQ at $-6$ dB are $49.79\%$ and $77.71\%$, respectively, while in terms of the mAP, the gaps are $67.93\%$ and $139.17\%$, respectively. The gain is attributed to the precise evaluation of the feature distortion in SemHARQ, where only distorted features are allowed to be retransmitted, improving the efficiency of channel resource utilization significantly, \textcolor{black}{especially for the low SNR regimes}.} 

\begin{figure*}[tbp]
	\centering
	\subfloat[{Color accuracy of different methods}]
  {\label{fig:a}\includegraphics[width=6cm]{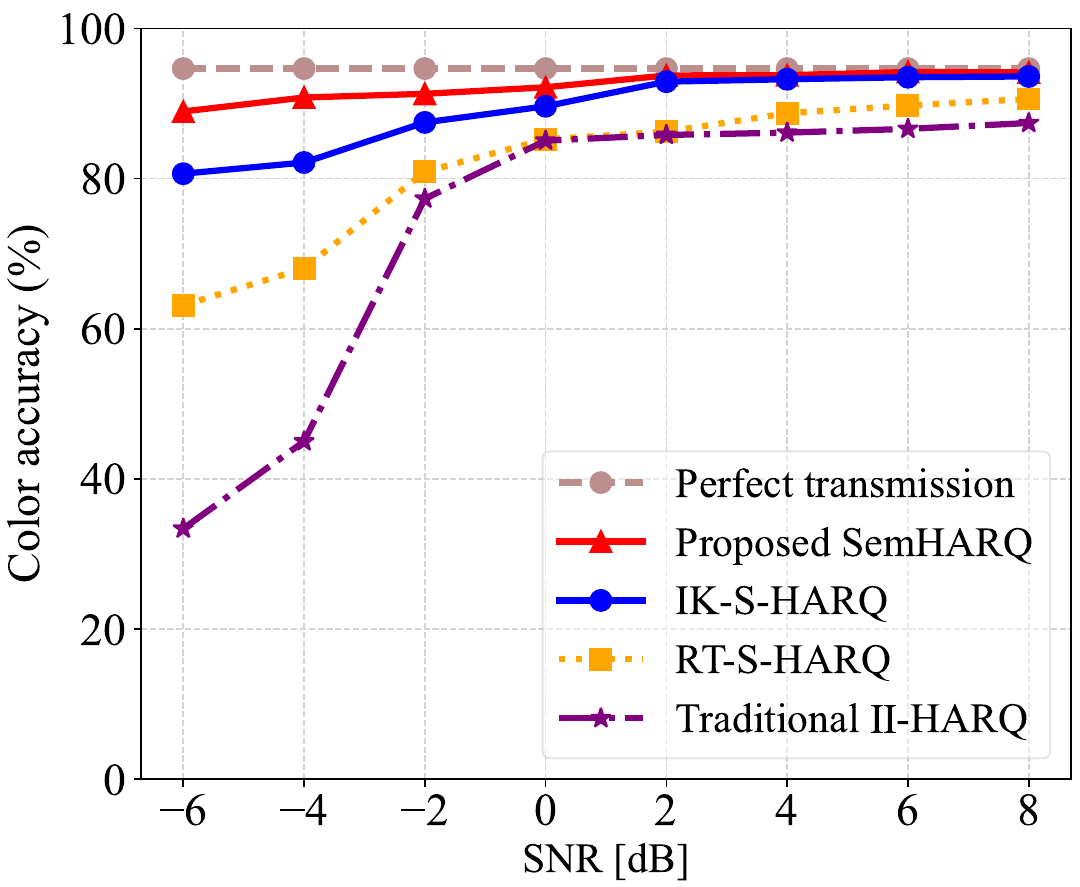}}\label{Fig_rank}\quad
   \vspace{1mm}
	\subfloat[{Type accuracy of different methods}]{\label{fig:b}\includegraphics[width=6cm]{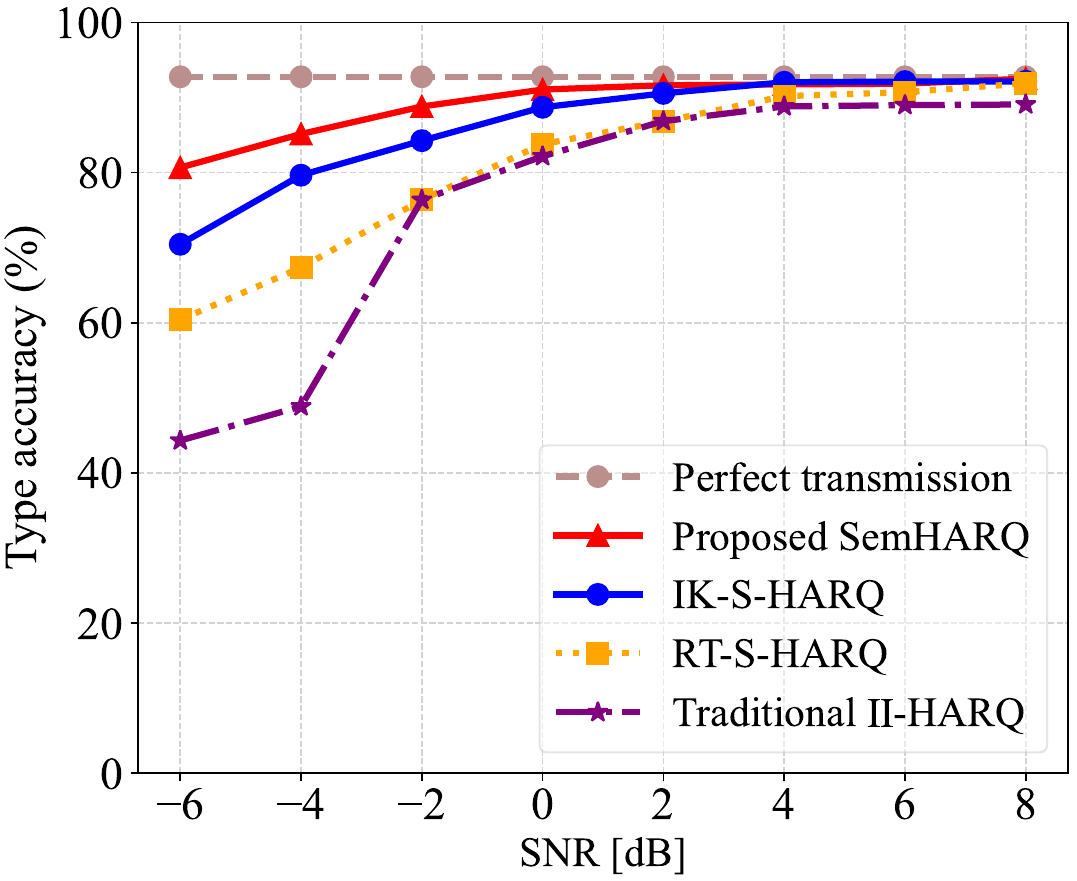}}\label{Fig_map}\\	
	\caption{  {Accuracy of vehicle color classification and type classification w.r.t. different methods.}}
	\label{Fig_classification}
\end{figure*}

\textcolor{black}{To further demonstrate the performance of SemHARQ at any given SNR regime, the task performance of the perfect transmission is shown in Fig.~\ref{Fig_vehicle_reid} and Fig.~\ref{Fig_classification}, which are obtained by performing the tasks directly using the extracted semantic feature vector without feature selection and transmission~(\textcolor{black}{i.e., transmission conditions are good enough to ensure perfect reconstruction of transmitted symbols at the receiver}, \textcolor{black}{as shown in the highest line in Fig.~\ref{Fig_vehicle_reid} and Fig.~\ref{Fig_classification}}). Note that to ensure a fair comparison, the same backbone DNN as SemHARQ is utilized. At low SNR regime, the transmission capacity is severely constrained and the transmitted features are heavily corrupted by the channel fading and noises. Therefore, under the constraint of the maximum number of retransmissions, the performance of SemHARQ is relatively poor. For example, at -6~dB, the SemHARQ exhibits a $70.97\%$ absolute reduction in rank-1 accuracy compared to the perfect transmission method in Fig.~\ref{Fig_vehicle_reid}.  However, this limitation is inherent to all HARQ methods operating under such transmission conditions. Particularly, the superiority of our SemHARQ lies in combining the retransmissions of distorted features with the incremental transmissions, which significantly improves the task performance compared to other HARQ methods, despite the capacity constraint. It is obvious that at high SNR regime, the performance of SemHARQ is very close to the perfect transmission, which results from the correct reception of most of the information under good channel conditions and high transmission capacity. For example, as presented in Fig.~\ref{Fig_vehicle_reid}, the rank-1 accuracy of the SemHARQ only reduces $0.44\%$  compared to the perfect transmission when SNR is $8$ dB. Furthermore, note that the performance gains of SemHARQ over IK-S-HARQ and RT-S-HARQ are also marginal when SNR is high. The reason is that the decoding is mostly successful after the first transmission, and thus, retransmission seldomly triggers under this condition. As a result, the superiority of SemHARQ over other semantic-aware HARQ methods degrades. }

At the same time, the accuracy of the other two classification tasks is illustrated in Fig.~\ref{Fig_classification}. The performance of SemHARQ is superior to other methods, \textcolor{black}{especially for the low SNR regimes}. Specifically, at $-6$ dB, in terms of color classification, SemHARQ improves the performance by $10.28\%$, $40.88\%$, and $166.48\%$, compared with the IK-S-HARQ, the RT-S-HARQ and the traditional HARQ, respectively. For type classification, the performance is improved by $14.50\%$, $33.55\%$, and $82.04\%$, respectively. {Note that since the classification task is less difficult than the ReID task, the accuracy is not affected by a certain amount of semantic errors. Therefore, the  semantic baselines perform well even without feature distortion evaluation, leading to less performance gain.
Additionally, the performance gain progressively reduces as SNR increases due to the correct decoding of the received semantics under good channel conditions. On this basis, the retransmissions become less meaningful.}

\begin{table}[t]
\caption{Performance comparison of the proposed FIR method and the SST method.}
\large
\scriptsize
\centering
\tabcolsep=0.1cm
\renewcommand\arraystretch{1.3}
 %\resizebox{1.2\columnwidth}{!}
 {
\begin{tabular}{cccccccccc}
\toprule
\multirow{2}{*}{{SNR~[dB]}} & \multicolumn{3}{c}{Proposed FIR ($\%$)} & \multicolumn{3}{c}{SST ($\%$)}  \\
\cmidrule(r){2-4} \cmidrule(r){5-7}
&  Rank-1 acc.       &  Color acc.  &  Type acc.
&  Rank-1 acc.     &  Color acc.  &   Type acc.  \\
\midrule
$-6$             &20.97        & 88.95    & 80.68                  & 1.61             & 42.54           & 47.51                   \\

$-4$             &23.65       & 90.81      & 85.15                   & 5.96           &70.76          &  70.54                   \\

$-2 $             &68.22       & 91.30    & 88.81                  &  27.83          & 82.24           &  82.15                    \\

$0$             &76.94    & 92.17      & 91.07                    & 68.00           & 90.72          & 87.67                      \\

$2$             &85.74        & 93.75        & 91.64                    & 83.91          & 92.67          & 90.97                   \\

$4$             &89.94          & 93.78     & 91.75                   & 88.97           & 92.80          & 91.45                     \\

$6$             &90.58     & 94.21        & 91.85                    & 89.52            & 93.47         & 91.29                      \\

$8$             &91.54       & 94.15       & 92.52                    & 90.01            & 93.54          & 91.54                     \\
\bottomrule
\end{tabular}}

\label{fir table}
\end{table}

\begin{figure*}[tbp]
	\centering
	\subfloat[{Rank-1 accuracy}]
  {\label{fig:a}\includegraphics[width=5.4cm]{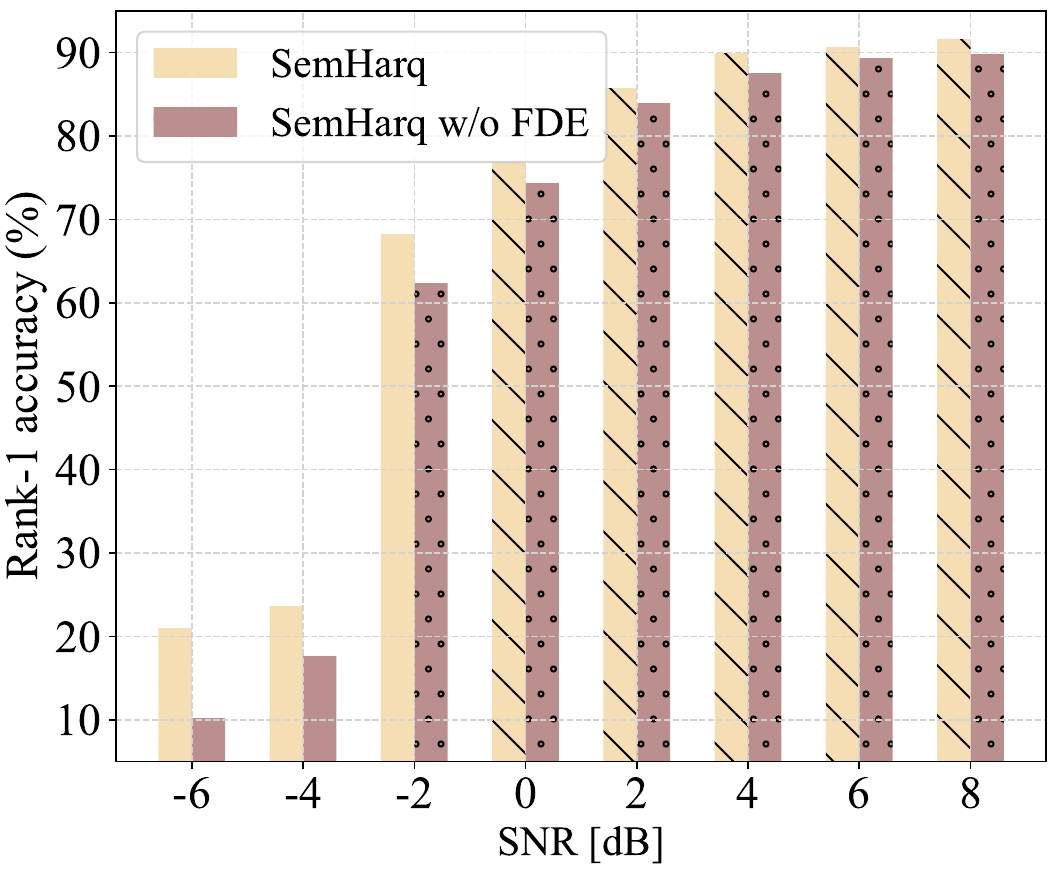}}\label{fde_rank1}\quad
   \vspace{1mm}
	\subfloat[{Color accuracy}]
 {\label{fig:b}\includegraphics[width=5.5cm]{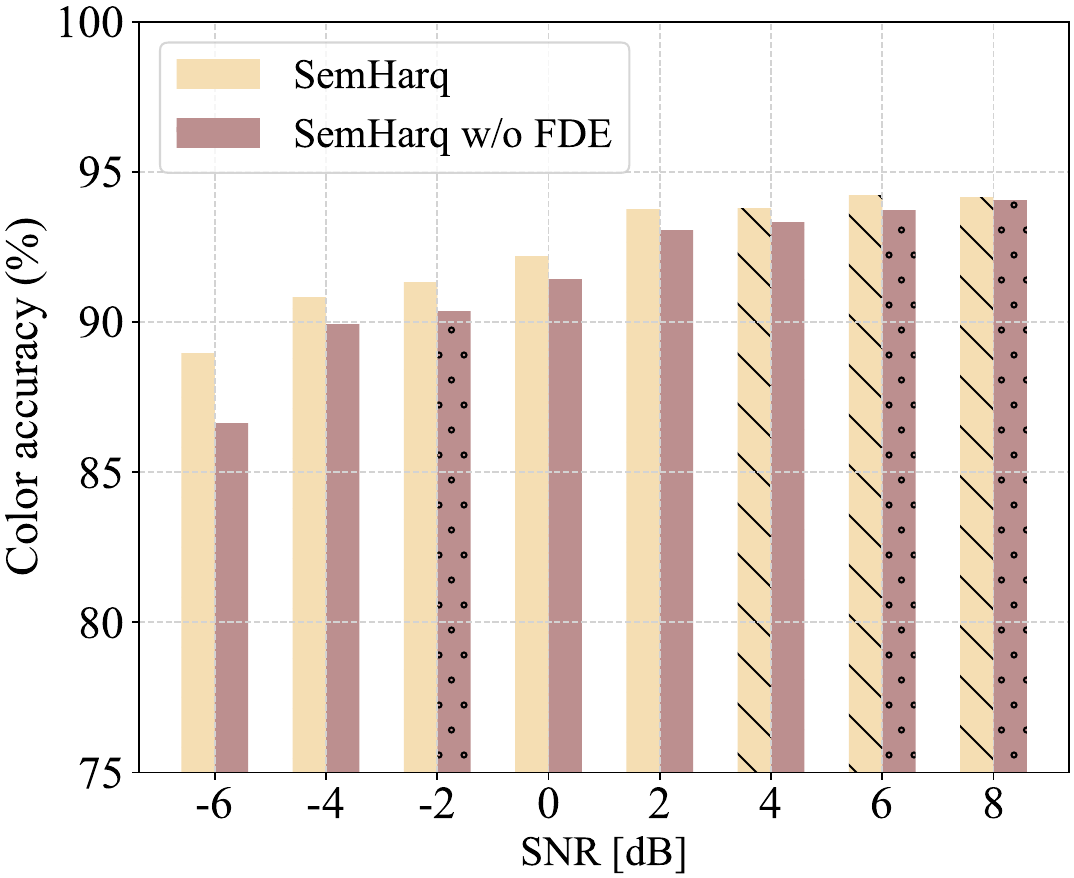}}\label{fde_color}
  \vspace{1mm}
	\subfloat[{Type accuracy}]
 {\label{fig:b}\includegraphics[width=5.5cm]{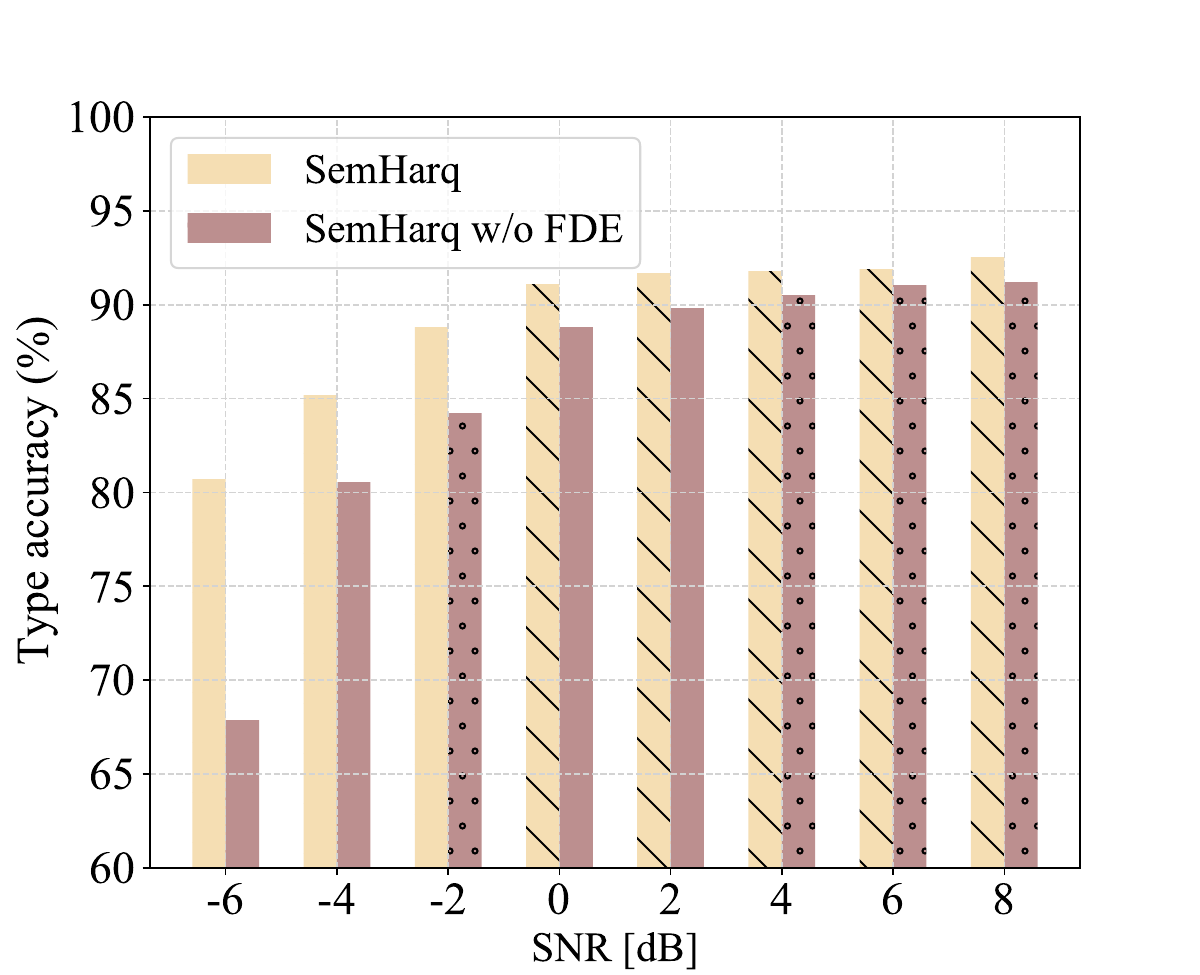}}\label{fde_type}
	\caption{ {Multi-task accuracy comparison between SemHARQ and SemHARQ without FDE, w.r.t. SNR.}}
	\label{without fde}
\end{figure*}

To further verify the effectiveness of the FDE network in SemHARQ, we conduct an ablation study for comparison. The distorted features are directly identified from the received feature vector ${\bm{\hat{z}}}^j$  using multi-dimension Gumbel-softmax in the comparison method. Accordingly, $\mathcal{F}$ and mutual information estimation discussed in \ref{FDE} are not involved, and FDE network training is not taken into account. As shown in Fig.~\ref{without fde}, the multi-task accuracy w.r.t. different channel conditions is demonstrated. It shows that our proposed SemHARQ achieves better task performance, especially with poor \text{SNR}s. Specifically, at $-6$ dB, the multi-task performance has been improved by $107.00\%$, $2.69\%$, and $18.91\%$ by using the FDE network, respectively. The reason is that the FDE network enhances the distortion evaluation ability of the system by measuring the contribution of the received semantics to the mutual information at each transmission. The obtained distortion levels quantify the degree of feature distortion, providing precise guidance on retransmission. Therefore, the efficiency of retransmission is improved dramatically.

The impact of FIR is also presented in Table~\ref{fir table}. The task performance is compared with the sequentially selected transmission (SST) method which transmits the first $B$ features without considering the importance levels at each transmission. Both FIR and SST perform scalable transmission w.r.t. various SNR levels. It is obvious that the FIR method surpasses the SST method significantly for all three tasks. Particularly, at $-2$ dB, the recognition accuracy increases by $145.13\%$, $11.02\%$, and $8.11\%$ in the three tasks, respectively. Note that vehicle ReID is more challenging due to the high intra-class variability coupled with small inter-class variability. However, greater advancements are made on the vehicle ReID than the remaining two tasks, demonstrating the superiority of the proposed system. One of the reasons is that the proposed FIR can build the relationship between the importance distribution of the semantic information and the multi-task predictions, thus effectively finding the optimal feature selection policy for partial semantics transmissions.  

\begin{figure}[t]
  \centering
\centerline{\includegraphics[width=6.5cm]{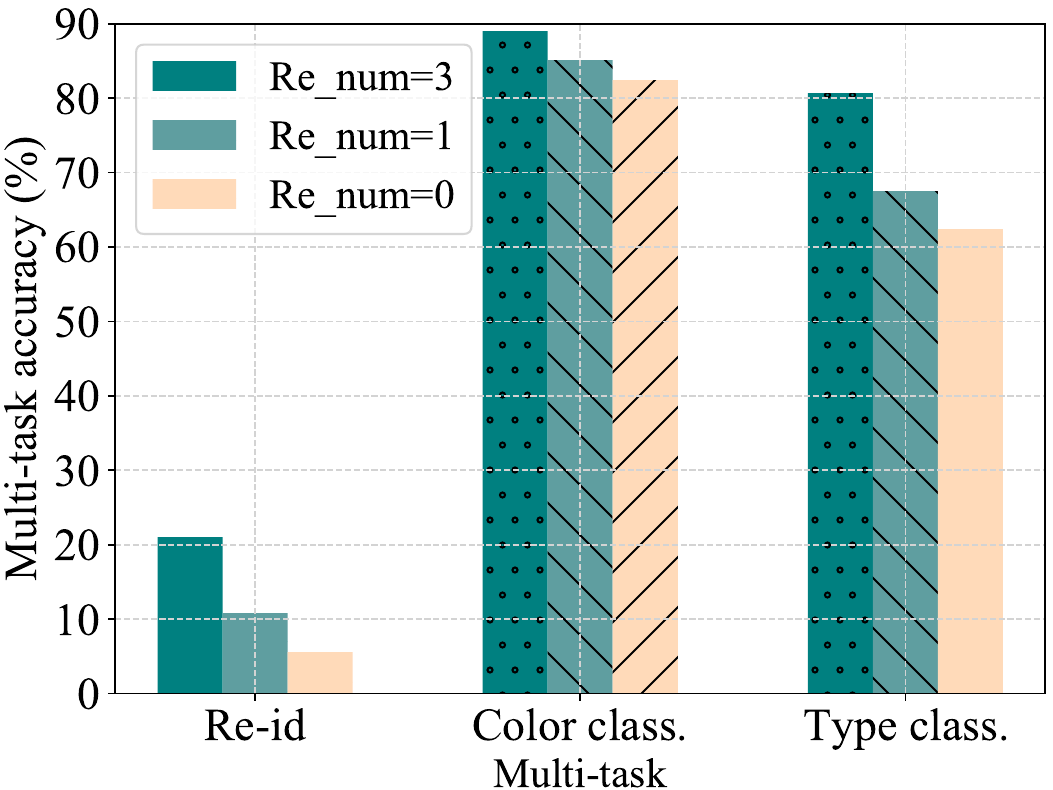}}
%  \vspace{2.0cm}
\caption{The impact of retransmissions on the accuracy of multi-task.}
\captionsetup[figure]{labelsep=space}
\label{re_num}
\end{figure}

Besides, to validate the effectiveness of the proposed SemHARQ more comprehensively, the gain of the system is investigated by increasing the number of retransmissions progressively. Fig.~\ref{re_num} presents the impact of retransmissions on the task performance in the case of $\text{SNR}=-6$ dB. Note that as the number of retransmissions increases, the system performance improves obviously. Particularly, the performance of vehicle ReID after three retransmissions is almost four times that of the no-retransmission method, which further proves the validity of SemHARQ.

\begin{figure}[t]
  \centering
\centerline{\includegraphics[width=6.2cm]{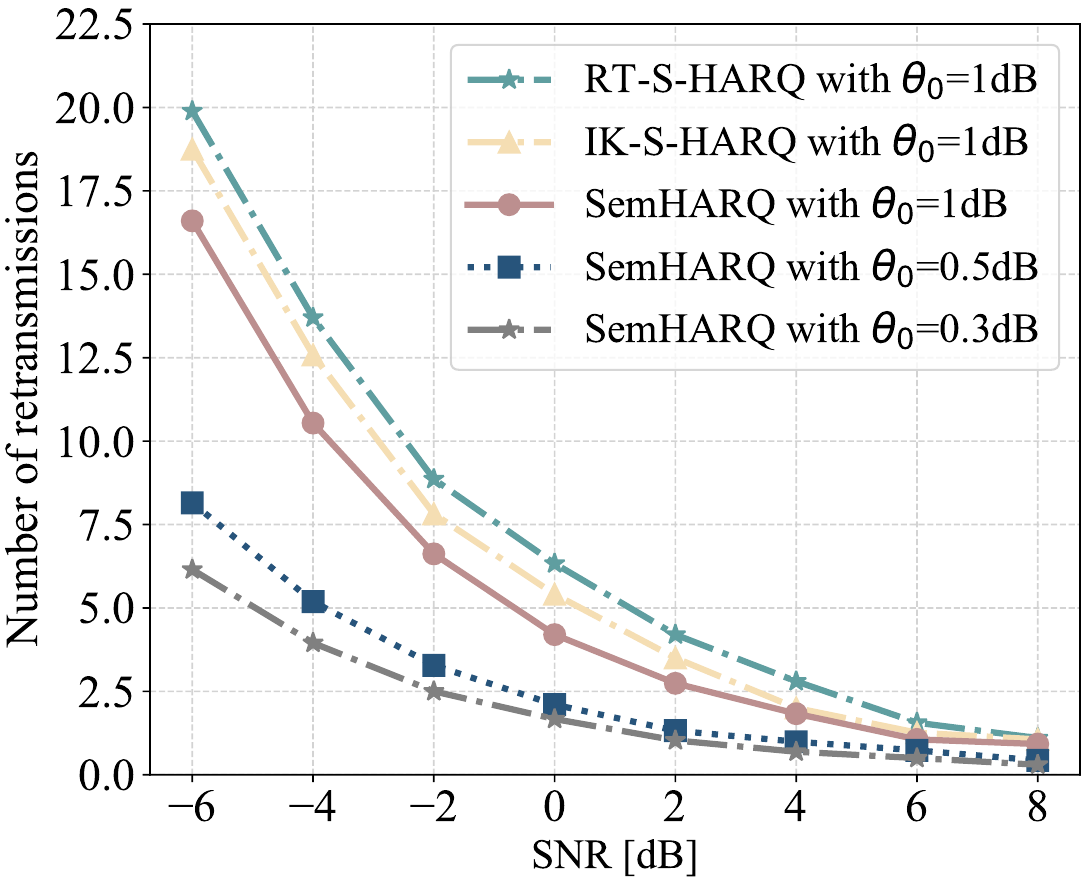}}
%  \vspace{2.0cm}
\caption{{Average number of retransmissions after using the retransmission identification module.}}
\captionsetup[figure]{labelsep=space}
\label{criteria}
\end{figure}

Moreover, to confirm the effectiveness of the retransmission identification module, the average numbers of retransmissions are summarized in Fig.~\ref{criteria} w.r.t. multiple $\text{SNR}$ levels, where the maximum number is set large enough. As shown in Fig.~\ref{criteria}, the retransmissions are more frequent under harsh channel conditions to meet the criteria in Eq.~(\ref{criteria_equa}). The average number of retransmissions at $-6$ dB is almost 18 times higher than at $8$ dB in SemHARQ when $\theta_0$, the
given conversion ratio between the uncertainty and SNR, is set to $1$ dB. Meanwhile, the need for retransmissions reduces when $\theta_0$ becomes lower accordingly, which represents that the receiver's demand for necessary information reception has been decreased. {Furthermore, note that the RT-S-HARQ and IK-S-HARQ methods require more frequent retransmissions than SemHARQ. Particularly, compared to SemHARQ, the numbers of retransmissions for RT-S-HARQ and IK-S-HARQ increase by $19.82\%$ and $12.95\%$ when SNR is set to $-6$ dB, respectively. The reason is that SemHARQ has higher resource utilization when combining the retransmission of distorted features and the incremental transmission of new features, reducing the need for retransmission.}

\begin{table}[t]
\centering
\caption{{Computational complexity and number of parameters of each module in SemHARQ.}}
\renewcommand{\arraystretch}{1.1} % 调整行高
\begin{tabular}{|>{\centering\arraybackslash}m{2.4cm}|>{\centering\arraybackslash}m{2.4cm}|>{\centering\arraybackslash}m{2.4cm}|}
\hline
\textbf{Transmission system} & \textbf{Complexity} & \textbf{Parameters} \\ \hline
Multi-task semantic encoder & $O(W \cdot H \cdot N)$ & $6.95~\text{M} $ \\ \hline
JSC-Encoder & $O(N^2)$  & $2.10~\text{M}$ \\ \hline
JSC-Decoder & $O(J \cdot B \cdot N)$  & $17.83~\text{M}$ \\ \hline
Semantic decoders and task performers & $O\big(N \cdot (M_1+M_2+M_3)\big)$  & $3.68~\text{M}$ \\ \hline
FDE network & $O(N^2)$  & $1.49~\text{M}$ \\ \hline
\end{tabular}
\label{table:complexity}
\end{table}

\begin{table}[t]
\centering
\caption{\textcolor{black}{Execution delay of each module in SemHARQ with batch size 100.}}
\renewcommand{\arraystretch}{1.1} % 调整行高
\begin{tabular}{|>{\centering\arraybackslash}m{5cm}|>{\centering\arraybackslash}m{2.8cm}|}
\hline
\textbf{Transmission system} & \textbf{Execution delay} \\ \hline
Multi-task semantic encoder & $55.27~\text{ms} $ \\ \hline
JSC-Encoder & $0.13~\text{ms}$ \\ \hline
Retransmissions & $2.2~\text{ms}$ \\ \hline
JSC-Decoder  & $0.14~\text{ms}$ \\ \hline
Semantic decoders and task performers & $0.27~\text{ms}$ \\ \hline
FDE network & $0.16~\text{ms}$ \\ \hline
\textcolor{black}{Overall inference latency} & $\textcolor{black}{58.17~\text{ms}}$ \\ \hline
\end{tabular}
\label{delay}
\end{table}

% \begin{table}[t]
% \caption{{Classification accuracy of different semantic-aware HARQ methods on ImageNet.}}
% \large
% \scriptsize
% \centering
% \tabcolsep=0.4cm
% \renewcommand\arraystretch{1.5}
% {
% \begin{tabular}{lccccccccc}
% \toprule
% %  & \multicolumn{9}{c}{Semantic-aware HARQ methods} \\
% % \cmidrule(lr){2-10}
% {SNR~[dB]}& \multicolumn{1}{c}{ SemHARQ~(Ours)} & \multicolumn{1}{c}{IK-S-HARQ} & \multicolumn{1}{c}{RT-S-HARQ} \\
% \midrule
% $-6$   & 21.36 & 15.85 & 12.16  \\
% $-2$   & 69.22  & 58.32 & 54.68 \\
% $2$    & 84.53  & 83.11 & 78.54 \\
% $6$    & 91.56  & 90.98 & 90.36 \\
% \bottomrule
% \end{tabular}}
% \label{imagenet}
% \end{table}

{\subsection{Discussions}}
\subsubsection{{Computational and implementation complexity}}
{ The computational complexity and number of parameters of different modules of SemHARQ are  shown in Table~\ref{table:complexity}. Specifically, the complexity of the multi-task semantic encoder is mainly impacted by the input and output dimensions of the DNN, 
where $W$, $H$ are the width and height of the input images, respectively, and $N$ denotes the size of the encoded semantic feature vector~\cite{IB-harq}. Compared with the JSC-Encoder, whose complexity is only dependent on the length of the semantic vector, the JSC-Decoder becomes more complicated as the total number of retransmissions $J$ increases. The reason is that the received feature vectors after all the transmissions are concatenated and decoded 
to the original length $N$ by the JSC-Decoder. Note that $B$ represents the dimension of the transmitted vector after scalable feature selection. Moreover, the complexity of semantic decoders and task performers is given by $O\big(N \cdot (M_1+M_2+M_3)\big)$~\cite{multi-task2}, where $M_1, M_2,$ and $M_3$ indicate the number of categories of the considered three tasks, respectively. It is obvious that more complex tasks require more complicated semantic decoders and task performers. Finally, the FDE network consists of multiple FC layers, whose complexity is determined by the input dimension $N$.
}

{Furthermore, as shown in Table~\ref{table:complexity}, the JSC-Decoder has the largest number of parameters due to the multiple transmissions. Meanwhile, since the DNN is used to extract semantic features, the parameter count of the multi-task semantic encoder is $6.95$~M, approximately one-third of that of the JSC-Decoder. Other modules have fewer parameters because they only consist  of FC layers with low input dimensions. Note that compared with the semantic-aware HARQ baselines, the only complexity increase of our proposed SemHARQ lies in the FDE network, which is negligible and acceptable. }

\textcolor{black}{Finally, an evaluation of the execution delay is conducted for different modules in the proposed SemHARQ system in Table~\ref{delay}. The delay is measured in terms of average time taken for each module to process a batch of data at an NVIDIA V100 GPU, where the batch size is set to 100. As shown in Table~\ref{delay},
the execution delay of the multi-task semantic encoder is the highest, approximately 300 times that of the other modules. The reason is that the multi-task semantic encoder is composed of DenseNet-121, which has 121 layers of convolutional neural network, resulting in relatively large computational cost.  However, the other modules like JSC-Encoder/Decoder, semantic decoders, task performers, each consists of two fully connected layers, respectively, while the FDE network comprises three, which is very lightweight with low execution delay. Additionally, the average retransmission delay under different SNRs is $2.2~\text{ms}$ with batch size 100, which is acceptable in practical systems.}

\begin{table}[t]
\caption{\textcolor{black}{\textcolor{black}{Classification accuracy of different semantic-aware HARQ methods on \texttt{ImageNet}}.}}
\large
\scriptsize
\centering
\tabcolsep=0.12cm
\renewcommand\arraystretch{1.0}
{
\begin{tabular}{lccccccccc}
\toprule
%  & \multicolumn{9}{c}{Semantic-aware HARQ methods} \\
% \cmidrule(lr){2-10}
{SNR~[dB]}& \multicolumn{1}{c}{ SemHARQ~(Ours)} & \multicolumn{1}{c}{IK-S-HARQ} & \multicolumn{1}{c}{RT-S-HARQ} \\
\midrule
$-6$   & \textbf{21.36} & 15.85 & 12.16  \\
$-2$   & \textbf{69.22}  & 58.32 & 54.68 \\
$2$    & \textbf{84.53}  & 83.11 & 78.54 \\
$6$    & \textbf{91.56}  & 90.98 & 90.36 \\
\bottomrule
\end{tabular}}
\label{imagenet}
\end{table}

\begin{figure}[t]
\centering
\centerline{\includegraphics[width=0.3\textwidth]{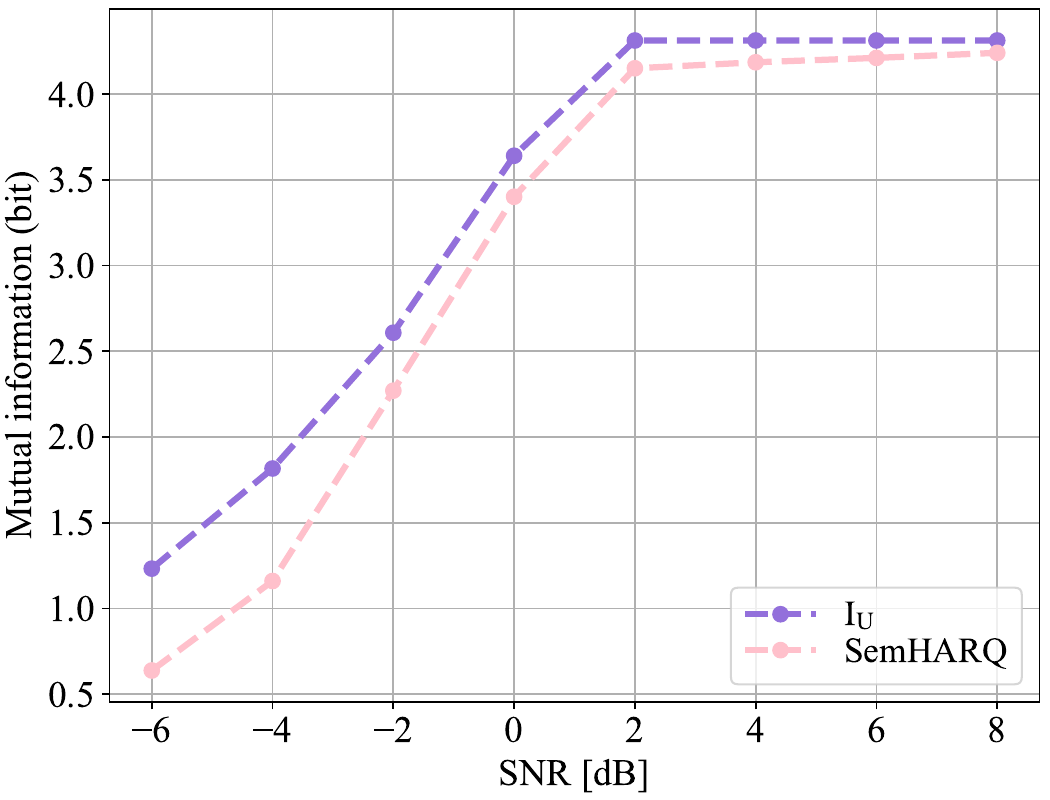}}
%\captionsetup{justification=centering}
\caption{\textcolor{black}{The mutual information of the upper bound and the proposed SemHARQ w.r.t. different SNRs.}}
%\captionsetup[figure]{labelsep=space}
\label{mi-esti}
\end{figure}

\subsubsection{{FDE module}}
\textcolor{black}{\textcolor{black}{Note that the FDE network - based on MINE - is the key module in the proposed SemHARQ.} As analyzed in \cite{mine2,mine1,mi}, when dealing with noisy variables with low relevance, \textcolor{black}{MINE may suffer from high variance}, particularly when handling high-dimensional inputs with insufficient batch size of training samples. For example, \textcolor{black}{for noisy vectors manifesting Gaussian distributions}, variance in the estimation tends to increase as the true mutual information grows when the dimension of the data is about 100 with batch size less than 64~\cite{mine1,mine2}. However, the task-related semantic features in SemCom frameworks exhibit higher semantic relevance and less randomness compared to randomly generated variables. Therefore, the uncertainty in the mutual information estimation is significantly reduced, leading to a stable MINE module, as in~\cite{mine3}. Note that the major variance introduced in SemHARQ \textcolor{black}{lies in the channel fading and additive noise}. To mitigate the impact, the transmission loss over the channel is combined with the mutual information loss during the training of the MINE module, which improves the scalability of different channel states.} \textcolor{black}{Additionally, the proposed FDE network is task-specific in our framework, which is consistent with the existing task-oriented semantic communication methods~\cite{task-oriented-jsac, jscc3}. However, the FDE methodology is general and can be applied to new semantic tasks by retraining or transfer learning techniques~\cite{task-unkown}. For example, the pretrained FDE network can be efficiently adapted to new tasks via fine-tuning or domain adaptation using a few new labels~\cite{task-unkown}.}

To further verify the scalability of the proposed SemHARQ on more complicated scenarios, \textcolor{black}{the classification accuracy of different semantic-aware HARQ methods on the \textcolor{black}{\texttt{ImageNet}} dataset with rician fading channel is shown in Table~\ref{imagenet}}. \textcolor{black}{Note that {\texttt{ImageNet}} is one of the most comprehensive image datasets}, encompassing a larger number of categories than our IoV scenario, with 1,000 diverse categories and varying data distributions~\cite{imagenet}.  As presented in Table~\ref{imagenet}, our proposed SemHARQ achieves the best performance among three semantic-aware HARQ methods, followed by IK-S-HARQ and RT-S-HARQ, \textcolor{black}{especially for the low SNR regimes}. Specifically, the classification accuracy gaps between SemHARQ and IK-S-HARQ and RT-S-HARQ at $-6$ dB are $34.76\%$ and $75.66\%$, respectively. \textcolor{black}{The performance
gain is attributed to SemHARQ’s design in accurately evaluating the incurring feature distortion because of an accurate estimation by MINE}, demonstrating the practicality of MINE for semantic features on diverse datasets. 

\subsubsection{Bounds of SemHARQ}
\textcolor{black}{To bound the performance margin between the optimal scheme and the proposed SemHARQ, we first depict the performance upper bound of the optimal scheme using mutual information. Given the optimal method that noisy transmission is not considered, the maximum mutual information between the semantic encoded feature vector and the multi-task prediction vectors is defined as follows:
\begin{align}
\textcolor{black}{
I_\text{DL} = \sum_{k=1}^{K} \max_{\alpha, \beta, \alpha^{-1}, {\beta_{k}^{-1}}, {\phi_{k}}} I\left(\bm{s}'; \bm{c}'_k \right),
}
\label{ubmi}
\end{align}
where $K$ represents the number of tasks, $\alpha, \beta, \alpha^{-1}, {\beta_{k}^{-1}}, {\phi_{k}}$ are the parameter sets of the semantic encoder, JSC-Encoder, JSC-Decoder, and the semantic decoder and task performer for the $k$-th task, respectively. Particularly, $I\left(\bm{s}'; \bm{c}'_k \right)$ denotes the mutual information between the semantic encoded feature vector $\bm{s}'$ and the $k$-th task prediction vector $\bm{c}'_k$. $I_\text{DL}$ indicates the optimal performance of the adopted deep learning model from the perspective of information theory. On this basis, we can obtain the ideal mutual information achievable at any given SNR, where \textcolor{black}{Rician channel model is considered}. Note that to ensure a fair comparison with our proposed SemHARQ, retransmission scheme is also utilized when calculating the theoretical limit of the mutual information under different SNRs.  Specifically, \textcolor{black}{the upper bound of the mutual information between the transmitter and the receiver is the total channel capacity with multiple transmissions}, which can be represented as follows,
\begin{align}
I_\text{U} &= \min (\frac{I_\text{DL}}{C}, G_\text{max}) \times C= \min ({I_\text{DL}},G_\text{max} \times C), 
\label{ubmisnr}
\end{align}
where $C=\log_2 \left( 1 + \text{SNR} \cdot \left\|\bm{h}\right\|^{2} \right)$ is the channel capacity with the fading coefficient vector $\bm{h}$\textcolor{black}{, $G_\text{max}$ is the considered maximum number of transmissions,} and $\min (\frac{I_\text{DL}}{C}, G_\text{max})$ denotes the actual number of transmissions. This indicates that $I_\text{DL}$ is the maximum mutual information achievable in SemCom systems at any SNRs. Finally, the mutual information margin between the upper bound and the proposed SemHARQ can be formulated as
\begin{align}
\textcolor{black}{\Gamma} &= 
\textcolor{black}{I_\text{U}- \sum_{k=1}^{K} \max_{\alpha, \beta, \alpha^{-1}, {\beta_{k}^{-1}}, {\phi_{k}}} I\left(\bm{s}; \bm{c}_k \right),}
\label{margin}
\end{align}
\textcolor{black}{where $I\left(\bm{s}; \bm{c}_k \right)$ denotes the mutual information between the semantic feature vector $\bm{s}$ at the transmitter and the $k$-th task prediction vector in SemHARQ, $\Gamma$ represents the discrepancy between the proposed SemHARQ and the optimal performance the system can achieve with certain deep learning encoding/decoding functions.} To further illustrate the performance margin, Fig.~\ref{mi-esti} presents $I_\text{U}$ and the estimated mutual information in SemHARQ. It shows that at high SNR regime, \textcolor{black}{e.g., at $8$ dB}, the mutual information of SemHARQ is only $1.7\%$ lower than $I_\text{U}$, indicating the effectiveness of the designed system. While at $-6$ dB, the mutual information of SemHARQ is nearly $50\%$ of the upper bound. \textcolor{black}{The poor performance may be due to the constant-coding rate commonly considered for SemCom systems}, although the designed HARQ mechanism can fix the constant-coding rate problem for a certain extent. \textcolor{black}{The performance could be further improved with an SNR-adaptive coding system, which can be an interesting topic for further research and design.}}

\bigskip

\section{Conclusions}

In this paper, a novel semantic-aware HARQ framework SemHARQ is proposed for multi-task semantic communications. Specifically, the task-related semantics are extracted by a multi-task semantic encoder and scalably transmitted in descending order of importance levels. On this basis, to enable the SemHARQ and improve the semantic error correction of the system, the distortion level of each received feature is evaluated by a feature distortion evaluation~(FDE) network. The corrupted features are identified by the distortion measure vector and retransmitted by the transmitter. Meanwhile, the incremental transmission is triggered to fully utilize the channel resources. Compared with the existing semantic-aware and traditional HARQ methods, the proposed SemHARQ has improved the multi-task performance by $16.27\%$-$166.48\%$ at low SNR regime.

%\section{References}

\small{
\bibliographystyle{IEEEtran}
\bibliography{IEEEabrv,reference}}

\begin{IEEEbiography}[{\includegraphics[width=1in,height=1.25in,clip,keepaspectratio]{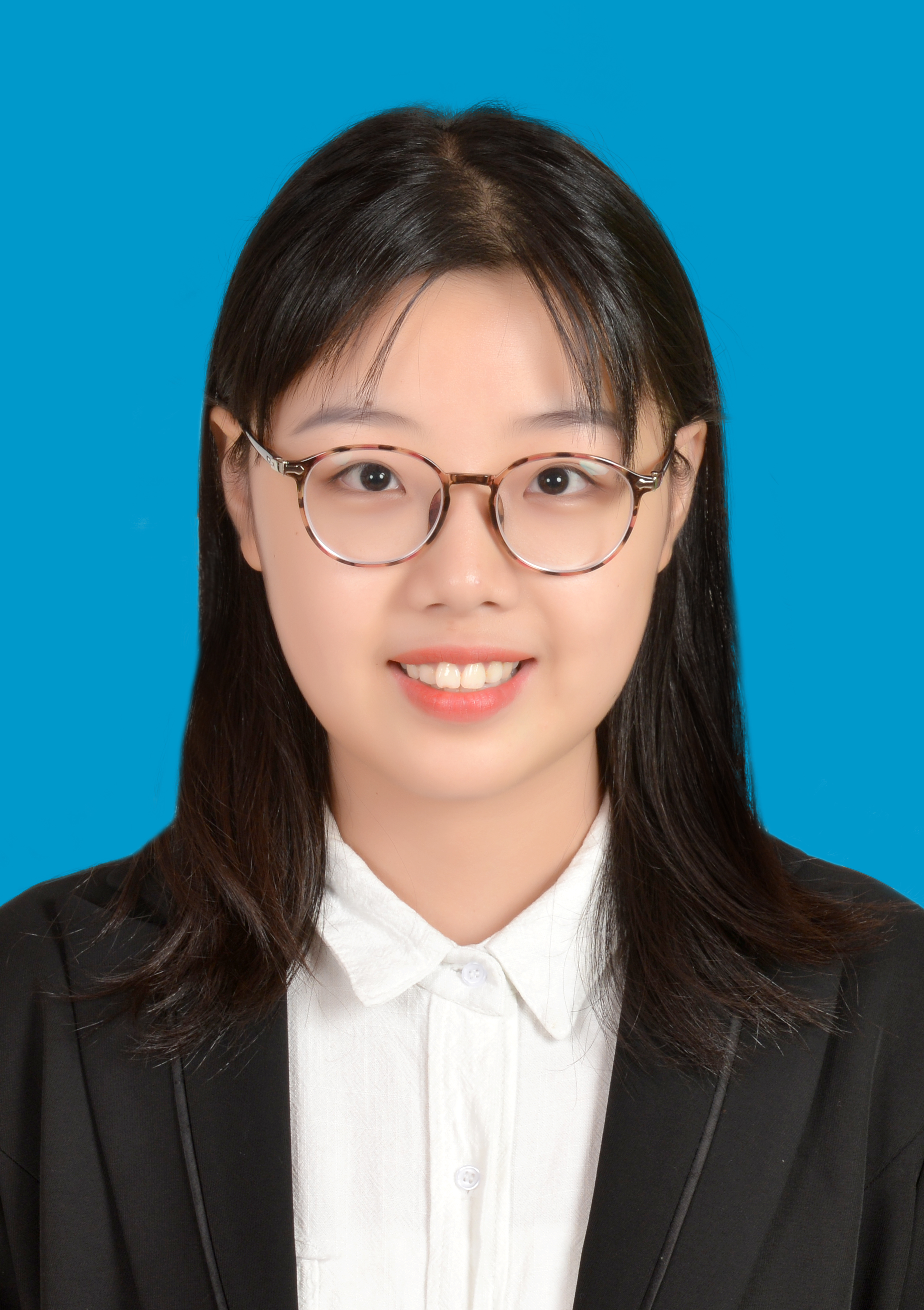}}]{Jiangjing Hu} is currently pursuing the Ph.D. degree with the School of Information and Communication Engineering, Beijing University of Posts and Telecommunications~(BUPT), Beijing, China.
Her research interests include semantic communications and image coding.
\end{IEEEbiography}

\begin{IEEEbiography}[{\includegraphics[width=1in,height=1.25in,clip,keepaspectratio]{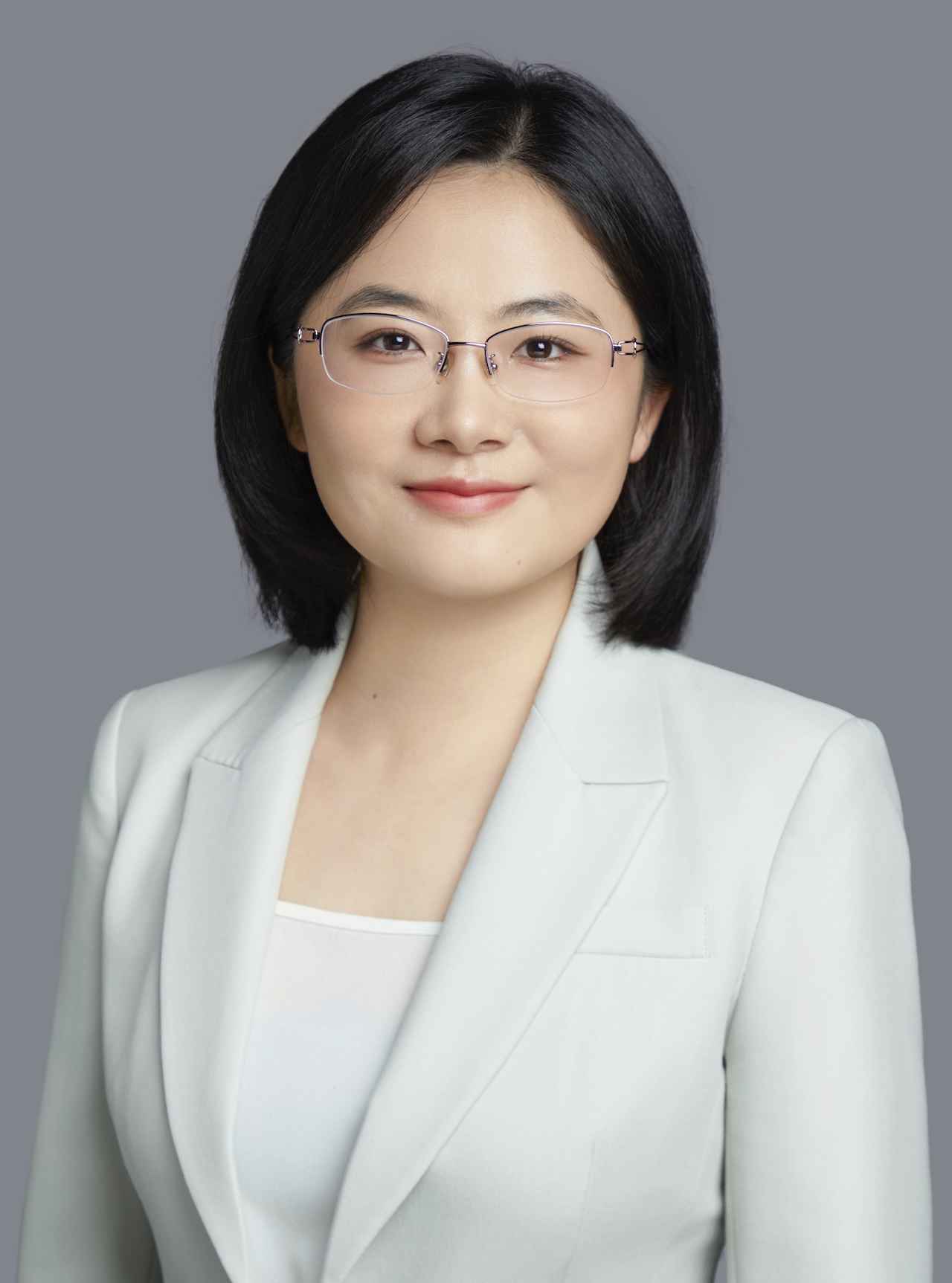}}]{Fengyu Wang} (Member, IEEE) received the B.S. and M.S. degrees from the School of Information and Communication Engineering, BUPT, Beijing, China, in 2014 and 2017, respectively, and the Ph.D. degree in electrical engineering from the University of Maryland at College Park (UMCP), College Park, MD, USA, in 2021.
She is currently a lecturer with the School of Artificial Intelligence, BUPT, Beijing, China.
Her research interests include semantic communications, wireless sensing, and statistical signal processing. She was the recipient of the Best Paper Award from IEEE GLOBECOM in 2022.
\end{IEEEbiography}

% \vspace{11pt}

\begin{IEEEbiography}[{\includegraphics[width=1in,height=1.25in,clip,keepaspectratio]{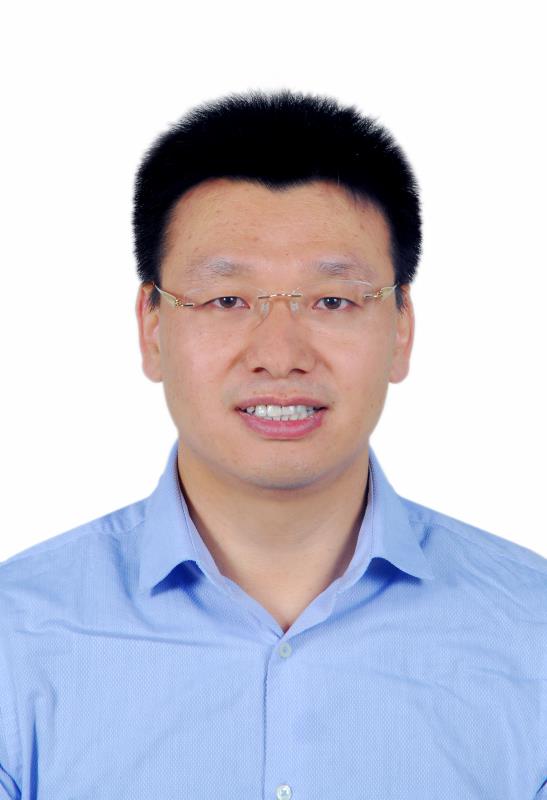}}]{Wenjun Xu} (Senior Member, IEEE) received the B.S. and Ph.D. degrees from BUPT, Beijing, China, in 2003 and 2008, respectively. He is currently a Professor and a Ph.D. Supervisor with the School of Artificial Intelligence, State Key Laboratory of Network and Switching Technology, BUPT, Beijing, China. He is currently an Editor of China Communications.
His research interests include AI-driven networks, semantic communications, UAV communications and networks, and green communications/networking. 
\end{IEEEbiography}

\begin{IEEEbiography}[{\includegraphics[width=1in,height=1.25in,clip,keepaspectratio]{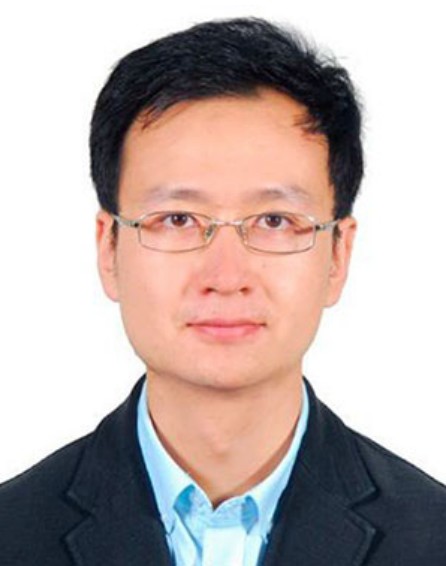}}]{Hui Gao} (Senior Member, IEEE) received the Ph.D. degree from the Beijing University of Posts and Telecommunications (BUPT), Beijing, China, in 2012. He is currently an Associate Professor with the School of Information and Communication Engineering, BUPT. His research interests include massive and mmWave MIMO systems, SAGINs, semantic communications, and intelligent wireless networks.
\end{IEEEbiography}

\begin{IEEEbiography}[{\includegraphics[width=1in,height=1.25in,clip,keepaspectratio]{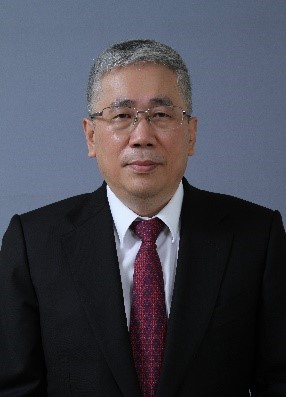}}]{Ping Zhang} (Fellow, IEEE) received the Ph.D. degree from BUPT, Beijing, China, in 1990, where he is currently a Professor. 
He is the director of State Key Laboratory of Networking and Switching Technology, a member of IMT-2020 (5G) Experts Panel, a member of Experts Panel for China’s 6G development. He served as Chief Scientist of National Basic Research Program (973 Program), an expert in Information Technology Division of National High-tech R\&D program (863 Program), and a member of Consultant Committee on International Cooperation of National Natural Science Foundation of China. His research interests mainly focus on wireless communication. He is an Academician of the Chinese Academy of Engineering (CAE).  
\end{IEEEbiography}

\end{document}